\pdfoutput=1

\documentclass[12pt]{article}
\usepackage{color}

\usepackage{graphicx}
\usepackage{cite,epsfig,bbm,amsmath,amssymb}

\def\be{\begin{equation}}
\def\ee{\end{equation}}
\def\bea{\begin{eqnarray}}
\def\eea{\end{eqnarray}}
\def\bal{\begin{align}}
\def\eal{\end{align}}

\def\slash#1{\setbox0=\hbox{$#1$}#1\hskip-\wd0\dimen0=5pt\advance
       \dimen0 by-\ht0\advance\dimen0 by\dp0\lower0.5\dimen0\hbox
         to\wd0{\hss\sl/\/\hss}}
\def\gev{{\rm GeV}}
\def\mev{{\rm MeV}}

\newcommand{\gae}{\lower 2pt \hbox{$\, \buildrel {\scriptstyle >}\over {\scriptstyle \sim}\,$}}
\newcommand{\lae}{\lower 2pt \hbox{$\, \buildrel {\scriptstyle <}\over {\scriptstyle \sim}\,$}}
\newcommand{\scs}{\scriptscriptstyle}
\newcommand{\f}{\frac}

\newcommand{\im}{$i$}
\begin{document}

\begin{titlepage}

\begin{flushright}
IUHET-545\\ [2cm]
\end{flushright}

\begin{center}

\setlength {\baselineskip}{0.3in} {\bf
An improved observable for the forward--backward asymmetry in $B\to K^* \ell^+ \ell^-$ and $B_s\to \phi \ell^+\ell^-$
}
\\[2cm]

\setlength {\baselineskip}{0.2in} {\large 
Enrico Lunghi$^1$ and Amarjit Soni$^{2}$}\\[5mm]

$^1$~{\it
Physics Department, Indiana University, Bloomington, IN 47405, USA\\
 E-mail: {\rm elunghi@indiana.edu}
}\\[5mm]
$^2$~{\it
 Physics Department, Brookhaven National Laboratory, \\
Upton, New York, 11973, USA\\
E-mail: {\rm soni@bnl.gov}
}\\[3cm]

\end{center}

\begin{abstract}
We study the decay $B\to K^* \ell^+ \ell^-$ in the QCD factorization approach and propose a new integrated observable whose dependence on the form factors is almost negligible, consequently the non--perturbative error is significantly reduced and indeed its overall theoretical error is dominated by perturbative scale uncertainties. The new observable we propose is the ratio between the integrated forward--backward asymmetry (FBA) in the $[4,6]\; {\rm GeV}^2$ and $[1,4]\; {\rm GeV}^2$ dilepton invariant mass bins. This new observable is particularly interesting because, when compared to the location of the zero of the FBA spectrum, it is experimentally easier to measure and its theoretical uncertainties are almost as small; moreover it displays a very strong dependence on the phase of the Wilson coefficient $C_{10}$ that is otherwise only accessible through complicated CP violating asymmetries. We illustrate the new physics sensitivity of this observable within the context of few extensions of the Standard Model (SM), namely the SM with four generations (SM4), an MSSM with non--vanishing source of flavor changing neutral currents in the down squark sector and a $Z^\prime$ model with tree level flavor changing couplings.
\end{abstract}

\setlength{\baselineskip}{0.2in}

\end{titlepage}

\section{Introduction}
\label{sec:introduction}
The importance of the flavor-changing semileptonic decays, $b \to s(d) l^+ l^-$ for searching effects of new physics has been recognized for a very long time \cite{HWS87,HN10,Browder:2008em,Egede:2010zc}.The process is especially important because its amplitude is very sensitive to the presence of heavy virtual quarks,  scaling as $({\rm mass}_{\rm quark})^2$. The inclusive ${\rm BR}(B \to X_s \ell^+ \ell^-) = (1.60 \pm 0.51) \times 10^{-6}$ is measured~\cite{Barberio:2008fa,Artuso:2009jw,Huber:2008ak} and is consistent with the NNLO prediction of the Standard Model (SM), $(1.62 \pm 0.14)\times 10^{-6}$ with an accuracy of around  9\%~\cite{Huber:2005ig,Huber:2007vv}; given the large experimental uncertainty, this measurement allows for new physics contributions at the $\sim 30\%$ level. The reaction is also of great  interest as it offers numerous related observables through which to test the SM precisely and to discover new phenomena. The dilepton invariant mass ($q^2$) spectrum, forward-backward asymmetry (FBA) as function of $q^2$, the location of the zero in  the FBA and the possibility of searching for CP violation via, for example, partial rate asymmetry (PRA) are some of the attractive features of this reaction. As usual while theory predictions tend to be more reliable for inclusive modes, experimentally the related exclusive modes such as $B\to K (K^*) \ell^+ \ell^-$, $B_s \to \phi \ell^+ \ell^-$ are more readily accessible.  A particular challenge for theory regarding these exclusive modes\cite{ALGH01} is their dependence on form factors, which are manifestly properties of bound states and therefore of non-perturbative character. B factories of course have been able to study both inclusive and exclusive modes but at the LHCb inclusive studies seem rather difficult. On the other hand many of the exclusive modes are quite distinctive and in particular $B\to K^* \ell^+\ell^-$, $B_s \to \phi \ell^+ \ell^-$ should be a focus of intense study at the LHCb and at the super--B factories.

In this paper, we will focus on the forward backward asymmetry in the $q^2$ distribution of $B\to K^* \ell^+ \ell^-$, $B_s \to \phi  \ell^+ \ell^-$. The topic has gained renewed interest recently as measurements at BaBar~\cite{:2008ju}, Belle~\cite{:2009zv} and CDF~\cite{cdf10047} all seem to show weak indications of deviations from expectations of the SM. Lack of statistics in these first observations may be the underlying reason but the prediction of the SM also has considerable uncertainties and that is a cause of some concern to which we will try to address in this work. As already mentioned  the dominant source of the uncertainties is the  form-factor dependence; these are not amenable to perturbation theory and improvement using non-perturbative methods tends to be rather slow.

Here we propose the ratio of the integrated forward--backward asymmetry in the $[4,6] \; \gev^2$ and $[1,4] \; \gev^2$ $q^2$ regions (just above and below the location of the zero of the spectrum). In this ratio, the form factors uncertainties cancel to a great extent and the total error is controlled by scale uncertainties whose perturbative origin allows for future improvements.

\section{$B\to K^* \ell \ell$ in QCD factorization}
\label{sec:qcdf}
The effective Hamiltonian responsible for the short distance $b\to s \ell^+ \ell^-$ transition is
\bal
{\cal H}_{\rm eff}^{b\to s \ell\ell} &= 
-\f{4 G_F}{\sqrt{2}} V_{tb}^{} V_{ts}^* \Bigg[\sum_{i=1}^{6} C_i O_i +\sum_{i=3}^6 C_{iQ} O_{iQ} + C_b O_b 
+ \sum_{i=7,8,S,P} ( C_i O_i + C_i^\prime O_i^\prime) \nonumber  \displaybreak[0]\\
& + \frac{\alpha_e}{4\pi} \sum_{i=9,10}  ( C_i O_i + C_i^\prime O_i^\prime) \Bigg]
\label{heff}
\end{align}
where the definition of the operators and the corresponding SM matching conditions can be found in Ref.~\cite{Huber:2005ig,Altmannshofer:2008dz}\footnote{Note that here we adopt a more conventional definition of the semileptonic operators $O_{9,10}$ and, unlike Ref.~\cite{Huber:2005ig}, we factor out $\alpha_e/4\pi$ in their definition so that their Wilson coefficients start at order $O(\alpha_s^0,\alpha_e^0)$.}. Primed operators are obtained from the unprimed ones by the substitution $L\leftrightarrow R$. The electroweak penguin and $O_b$ operators contributions to $B\to K^* \ell\ell$ can be neglected because their impact is much smaller than the overall theory uncertainty. 

It is possible to show that, for small values of the dilepton invariant mass ($q^2 \lae 6 \; \gev^2$), the $B\to K^* \ell\ell$\footnote[3]{For definiteness, we will continue to explicitly mention $B\to K^* \ell^+ \ell^-$ only, however, it should be obvious that our equations and the ensuing discussions are applicable equally well to $B_s \to \phi \ell^+\ell^-$.} matrix element of all the operators appearing in Eq.~(\ref{heff}) obey the following factorization formula~\cite{Beneke:2001at}:
\bea
\langle K^*_a \ell\ell | O_i | B \rangle = 
C_a^i \; \xi_a^{}(q^2) +  
\sum_\pm \int {\rm d}\omega \; {\rm d}u \; T_{a,\pm}^i (q^2,u,\omega)  \; \Phi_{B,\pm}^{} (\omega)\; \Phi_{K^*}^{} (u)  
+ O\left( \Lambda_{\scs QCD}/m_b\right)
\label{factorizationformula}
\eea
where $a= \perp,\parallel$ refers to the transverse (longitudinal) polarization of the $K^*$, $\xi_a$ are soft form factors and  $\Phi_{B,\pm}^{}$ and $\Phi_{K^*}^{}$ are the light cone wave functions of the $B$ and $K^*$ mesons. The quantities $C_a^i$ and $T_{a,\pm}^i$ can be calculated in perturbation theory. 

$B\to K^* \ell\ell$ matrix elements can be decomposed using a basis of eight transversity amplitudes (see, for instance, the discussion in Sec.~3.2 of Ref.~\cite{Altmannshofer:2008dz}): $A_{\perp L,R}$, $A_{\parallel L,R}$, $A_{0L,R}$, $A_t$ and $A_S$. In this notation, $A_\perp$ and $A_\parallel$ correspond to the two possible transverse polarization states of the $K^*$ and are described by the form factor $\xi_\perp$; $A_0$, $A_t$ and $A_S$ involve a longitudinally polarized $K^*$ and are described in terms of $\xi_\parallel$. 

Up to electromagnetic corrections, the various $B\to K^*\ell\ell$ matrix elements of the semileptonic and magnetic moment operators $O_{7,9,10}^{(\prime)}$\footnote[2]{This applies also to the matrix elements of the scalar and pseudoscalar operators $O_{S,P}^{(\prime)}$.} can be exactly expressed in terms of the seven form factors $V(q^2)$, $A_{1,2,3}(q^2)$ and  $T_{1,2,3}(q^2)$. Up to power corrections, these form factors can be written as the sum of soft and hard factorizable contributions using Eq.~(\ref{factorizationformula}). It is important to realize that the separation between soft and hard contributions is subject to a certain degree of arbitrariness; therefore it is necessary to choose a factorization scheme and adopt precise definitions of the soft form factors in terms of the full QCD ones. Following the analysis of Ref.~\cite{Beneke:2004dp} we adopt the following definitions:
\bea
\xi_\perp (q^2)= \f{m_B}{m_B+m_{K^*}} V(q^2) 
\quad {\rm and} \quad
\xi_\parallel(q^2) = \f{m_B + m_{K^*}}{2 E_{K^*}} A_1(q^2) -  \f{m_B - m_{K^*}}{m_B} A_2(q^2)
\; .
\label{factorizationscheme}
\eea
The form factors $V$, $A_1$ and $A_2$ are taken from the Light-Cone Sum Rule analysis of Refs.~\cite{Ball:2006nr,Altmannshofer:2008dz}. In the numerical analysis we vary independently the $q^2 = 0$ values of the two soft form factors $\xi_\perp$ and $\xi_\parallel$ over the one--sigma errors given in Table~\ref{tab:inputs}. The matrix elements of all the other operators do not admit a simple interpretation in terms of form factors but can be expressed using Eq.~(\ref{factorizationformula}) in which $\xi_\perp$ and $\xi_\parallel$ are the same soft form factors introduced in Eq.~(\ref{factorizationscheme}).

Let us now sketch how all the different parts of this calculation come together. Up to $O(\alpha_s)$ corrections, the three most relevant transversity amplitudes are schematically given by
\bea
A_{\perp L,R} & \propto & [(C_9^{\rm eff} +C_9^\prime) \mp (C_{10}^{} + C_{10}^\prime)] \; V(q^2) + \f{2 m_b m_B }{q^2} (C_7^{} + C_7^\prime) \;  T_1 (q^2) \label{aperp}\\
A_{\parallel L,R} & \propto & [(C_9^{\rm eff} -C_9^\prime) \mp (C_{10}^{} - C_{10}^\prime)] \;  A_1(q^2) + \f{2 m_b m_B }{q^2} (C_7^{} - C_7^\prime) \;  T_2 (q^2) \label{aparallel}\\
A_{0 L,R} & \propto & [(C_9^{\rm eff} -C_9^\prime) \mp (C_{10}^{} - C_{10}^\prime)] \; \{A_1(q^2), A_2 (q^2)\} + (C_7^{} - C_7^\prime)\; \{T_1(q^2) ,T_2 (q^2)\}
\eea
where $\{ \cdot  , \cdot \}$ indicates a linear combination, $C_9^{\rm eff} = C_9 + Y (q^2)$ and $Y(q^2)$ is the sum of the leading order matrix elements of the current--current and QCD penguin operators $O_{1-6}$ and can be found for instance in Ref.~\cite{Beneke:2001at}. All Wilson coefficients are evaluated at a scale $\mu_b \sim m_b$. Complete expressions for the eight transversity amplitudes can be found in Refs.~\cite{Kruger:2005ep, Lunghi:2006hc,Altmannshofer:2008dz}. 

Part of the $O(\alpha_s)$ corrections are buried inside the form factors; therefore the first step consists in replacing each form factor with the corresponding QCD factorization expansion (e.g. $V = \xi_\perp \; (m_B+m_{K^*})/m_B$). The complete expansion of the form factors in terms of $\xi_{\perp,\parallel}$ can be found in Refs.~\cite{Beneke:2000wa,Beneke:2001at,Beneke:2004dp}. In a second step all the remaining non-factorizable corrections originating from the remaining operators (other than $O_{7,9,10}$) have to be included. Since the latter can always be described in terms of $B\to K^* \gamma^*$ matrix elements it is customary to lump them together in some effective photonic form factors:
\bea
( C_7^{} - C_7^\prime  ) \; T_i (q^2)  &\rightarrow& \tau_i^- (q^2)\\
( C_7^{} + C_7^\prime  ) \;  T_i(q^2) &\rightarrow& \tau_i^+(q^2) \; ,
\eea
where the index $\pm$ refers to the combinations $C_7^{} \pm C_7^\prime$. The quantities $\tau_i^\pm$ are obtained from the $\tau_i$ introduced in Ref.~\cite{Beneke:2001at} with the appropriate $C_7 \rightarrow C_7^{} \pm C_7^\prime$ substitution. Following Ref.~\cite{Beneke:2001at}, the contribution from the $Y(q^2)$ term in $C_9^{\rm eff}$ is also included in the $\tau_i$; therefore we also have to replace $C_9^{\rm eff} \rightarrow C_9$. 

Note that the factorization scheme that we adopt implies the absence of $O(\alpha_s)$ corrections to the form factors $V(q^2)$, $A_1(q^2)$ and $A_2 (q^2)$ (that are therefore simply expressed in terms of the soft form factors $\xi_{\perp,\parallel}$ without any $O(\alpha_s)$ contribution). The form factor $A_0 (q^2)$, on the other hand, receives non-trivial $O(\alpha_s)$ corrections. Effects on the photonic form factors $T_i (q^2)$ are included in the definition of the $\tau_i^\pm$.

Once the transversity amplitudes have been calculated at order $\alpha_s$, it is trivial to express the fully differential $B\to K^* (\to K \pi) \ell\ell $ decay width. In the limit $m_\ell \to 0$, one finds:
\bea
\f{{\rm d} \Gamma}{ {\rm d} q^2 \; {\rm d} \cos\theta_\ell} & = & \f{3}{4} 
\Big(I_1^c  +2 I_2^s +  (2 I_2^s-I_1^c) \;\cos^2\theta_\ell+ I_6^s\; \cos \theta_\ell 
\Big)
\label{differentialwidth}
\eea
where 
\bea
I_2^s  &=& \f{1}{4} \left( | A_\perp^L|^2 + | A_\parallel^L|^2 + | A_\perp^R|^2 + | A_\parallel^R|^2\right)\\
I_1^c &=&  | A_0^L|^2 + | A_0^R|^2 \\
I_6^s & = & 2\; {\rm Re} \left(  A_\parallel^L A_\perp^{L*} -A_\parallel^R A_\perp^{R*}    \right)
\label{i6s}
\eea
where $\theta_\ell$ is the angle between the $\ell^+$ and the $B$ in the dilepton centre of mass system. Complete expressions for the differential decay width and the twelve $I_i^a$ functions can be found in Ref.~\cite{Altmannshofer:2008dz}. 
\begin{table}[t]
\begin{center} 
\begin{tabular}{|l|l|} \hline
\vphantom{$\Big($}$m_b^{1S} = (4.68 \pm 0.03 )\; \gev$~\cite{Bauer:2004ve}& 
$m_c(m_c) = (1.27^{+0.07}_{-0.11} ) \; {\rm GeV}$~\cite{pdg} \cr
$\xi_\perp (0) = 0.266 \pm 0.032$~\cite{Ball:2006nr,Altmannshofer:2008dz}&  
$\xi_\parallel (0) = 0.118 \pm 0.008$~\cite{Ball:2006nr,Altmannshofer:2008dz}\cr
$f_B = (192.8 \pm 9.9 )\; \mev$~\cite{Laiho:2009eu}  & 
$f_{k^*}^\perp = (163 \pm 8) \; \mev$~\cite{Ball:2006fz} \cr
$f_{k^*}^\parallel = (220 \pm 5) \; \mev$~\cite{Ball:2006fz}  &
$a_1^{\perp,\parallel} = 0.03 \pm 0.03$~\cite{Ball:2006fz} \cr
$a_2^{\perp,\parallel} = 0.08 \pm 0.06$~\cite{Ball:2006fz} &
$\lambda_B = (0.51 \pm 0.12)\; \gev$~\cite{Ball:2006nr}\cr
\hline
\end{tabular}
\caption{Inputs used in the numerical analysis of the forward--backward asymmetry. All other inputs are taken from the PDG~\cite{pdg}. \label{tab:inputs}}
\end{center}
\end{table}
The forward--backward asymmetry is obtained integrating Eq.~(\ref{differentialwidth}) over $\cos \theta_\ell$:
\bea
{\cal A}_{\rm FB} (q^2) & \equiv & 
\f{\int \f{{\rm d} \Gamma}{ {\rm d}q^2 \; {\rm d}\cos\theta_\ell} {\rm sgn} (\cos\theta_\ell)\; {\rm d} \cos\theta_\ell }{\int  \f{{\rm d} \Gamma}{ {\rm} d q^2 \; {\rm d}\cos\theta_\ell} \; {\rm d} \cos\theta_\ell  }
= \f{3}{4} \f{I_6^s}{I_1^c + 4 I_2^s} \; .
\label{afb}
\eea
Using Eqs.~(\ref{aperp},\ref{aparallel},\ref{i6s},\ref{afb}) and the leading order expressions for the form factors, one finds that the numerator of the forward--backward asymmetry is proportional to 
\bea
- \xi_\perp^2 {\rm Re}  \left\{ C_{10}^* \left( C_9^{\rm eff} (q^2)+2 \f{m_{b} m_B}{q^2} C_7 \right) \right\} + O(\alpha_s,\Lambda/m_b) \; .
\eea
At order $O(\alpha_s^0)$ this quantity does not depend on the form factor $\xi_\parallel$. For this reason hadronic uncertainties affect the location of zero of the ${\cal A}_{\rm FB}$ spectrum
\bea
q^2_0 &\simeq& -2 m_{b} m_B \f{C_9^{\rm eff} (q_0^2)}{C_7} + O(\alpha_s,\Lambda/m_b)
\eea
only at the NLO level. Note that this statement does not extend to integrals of the forward--backward asymmetry over an arbitrary $q^2$ interval. In fact the denominator of Eq.~(\ref{afb}), i.e. the differential rate, depends at leading order on both soft form factors through the transversity amplitude $A_{0L,R}$: at this order the form factors $V$, $A_0$, $T_1$ and $T_2$ are proportional to $\xi_\perp$ and the form factor $A_2$ is proportional to $\xi_\perp - \xi_\parallel$. The uncertainty on our knowledge of these two form factors is the dominant source of uncertainty on the forward--backward asymmetry spectrum and various integrals. In Table~\ref{tab:inputs} we summarize the inputs that we need for a complete numerical analysis of the forward--backward asymmetry; note that the $\xi_{\perp,\parallel}(0)$ are taken from Ref.~\cite{Ball:2006nr,Altmannshofer:2008dz} from where the $q^2$ dependence of the form factors is also taken.

\begin{figure}
\begin{center}
\includegraphics[width=0.47 \linewidth]{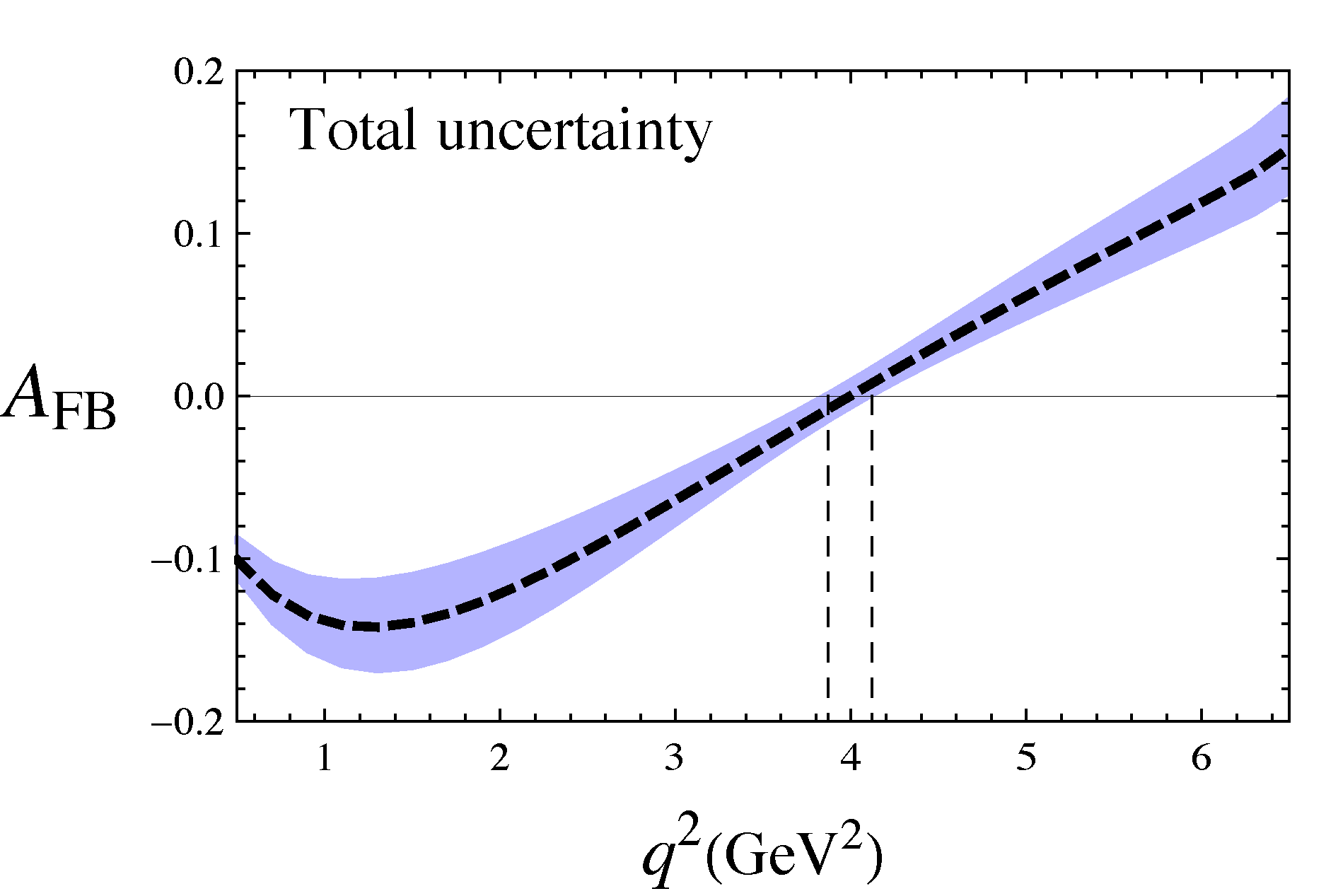}
\includegraphics[width=0.47 \linewidth]{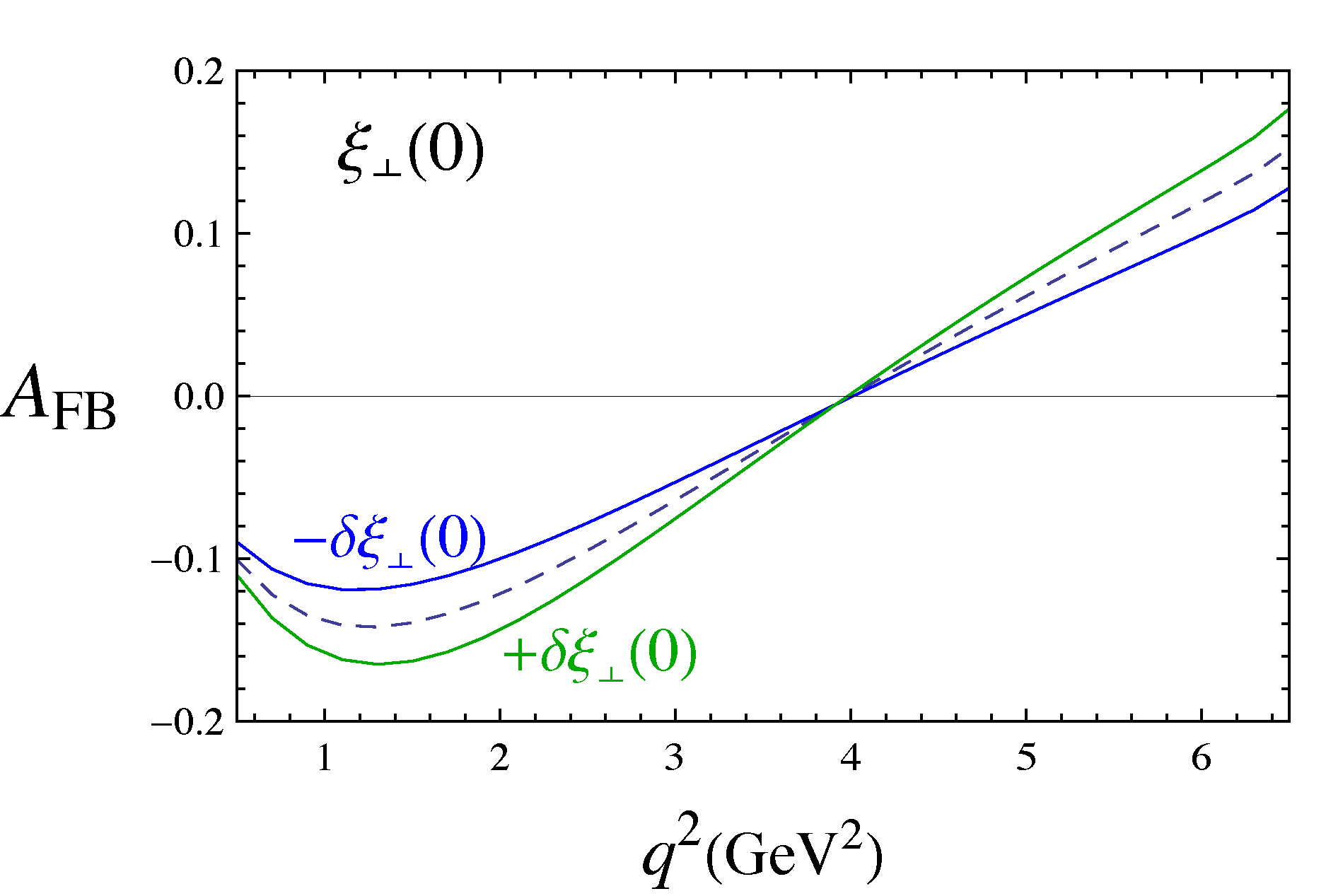}
\includegraphics[width=0.47 \linewidth]{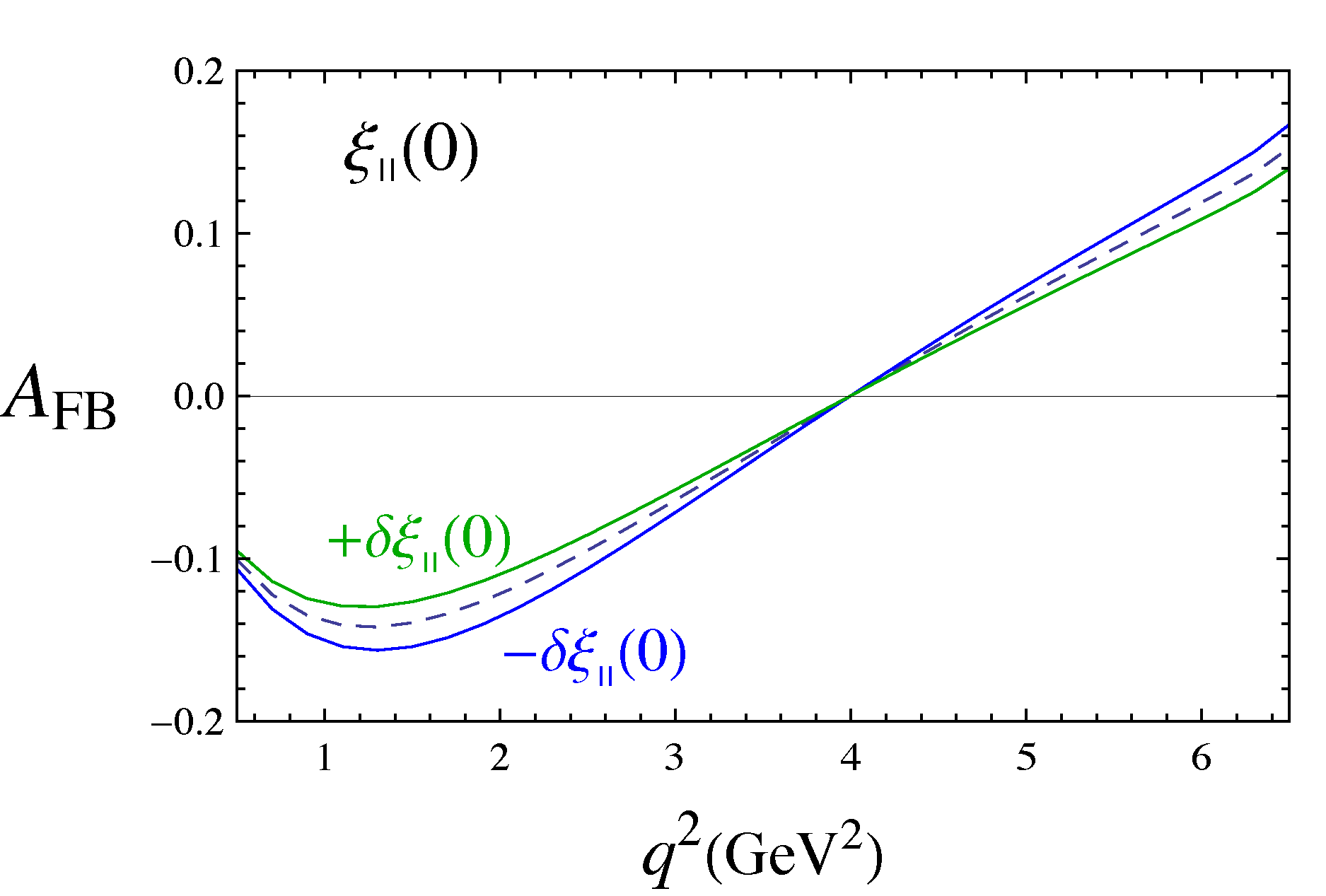}
\includegraphics[width=0.47 \linewidth]{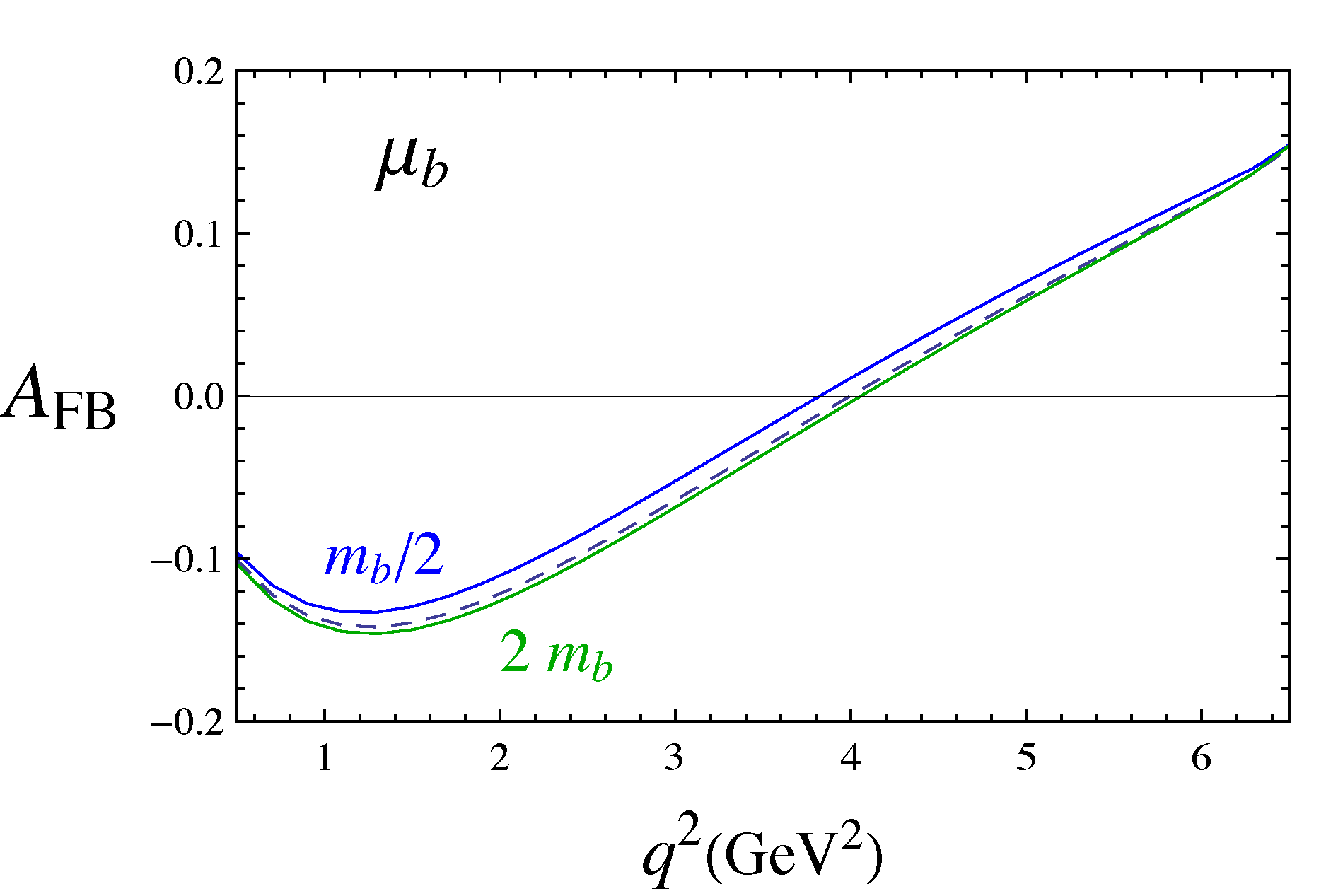}
\end{center}
\vskip -0.7cm
\caption{Error analysis of the forward--backward asymmetry spectrum. In the first panel we present the total error on the FBA spectrum obtained by adding all sources of uncertainty in quadrature. In the remaining panels we show the errors induced separately by the two soft form factors and by the variation of the low--scale $\mu_b$. \label{fig:AFB}}
\end{figure}

In Fig.~\ref{fig:AFB} we plot the differential forward--backward asymmetry in the $1\; \gev^2 < q^2 < 6 \; \gev^2$ range. The blue shaded area in the first panel indicates the total uncertainty that we obtain by varying all the input parameters given in table~\ref{tab:inputs} and the factorization scale $\mu_b$. The latter dependence is an artifact and can be reduced by including higher order terms of the perturbative $\alpha_s$ expansion. In this paper we allow $\mu_b$ to vary in the usual $[m_b^{\rm pole}/2, 2 m_b^{\rm pole}]$ range (but also present the results we obtain for the more conservative $[4,5.6]~{\rm GeV}$ range adopted in Ref.~\cite{Altmannshofer:2008dz}). The location of the zero of the forward--backward asymmetry spectrum is $q_0^2 = (4.0 \pm 0.13) \; \gev^2$. In the other three panels we show separately the impact of the three most important sources of uncertainties. Note that the variation of the scale $\mu_b$ at which the Wilson coefficients are evaluated induces a constant shift on the whole spectrum and that the uncertainties on the $q^2 = 0$ values of the soft form factors switch sign above and below $q_0^2$. 

The latter observation leads us to propose a new integrated observable for which we expect a very tiny form factor uncertainty, namely the ratio between the the forward--backward asymmetry integrated in the $[4,6]\; \gev^2$ and $[1,4]\; \gev^2$ bins:
\bea
{\cal R} [q_0^2] & = & \f{
\int_{q_0^2}^6 {\cal A}_{\rm FB} (q^2) \; {\rm d} q^2
}{
\int_1^{q_0^2} {\cal A}_{\rm FB} (q^2) \; {\rm d} q^2
} \; .
\eea
The SM predictions for the integrated forwards--backward asymmetry in the two low-$q^2$ bins and for the ratio ${\cal R}[q_0^2]$ are:
\bea
\int_1^4 {\cal A}_{\rm FB} (q^2) \; {\rm d} q^2 & = & -0.089\;  ( 
1 \pm 0.085_{\mu_b} \pm 0.17_{\xi_\perp(0)} \pm 0.10_{\xi_\parallel(0)} \pm 0.031_{\rm rest} ) \label{lowerSM1}\\
&=& -0.089 \pm 0.020\; [22\%]\; , \label{lowerSM}\\
\int_4^6 {\cal A}_{\rm FB} (q^2) \; {\rm d} q^2 & = &   0.062\;  ( 
1 \pm 0.089_{\mu_b} \pm 0.18_{\xi_\perp(0)} \pm 0.10_{\xi_\parallel(0)} \pm 0.067_{\rm rest} )  \label{higherSM1}\\
&=& 0.062 \pm 0.015\; [24\%]\; , \label{higherSM}\\
{\cal R} [4] & = & -0.701 \; (
1 \pm 0.19_{\mu_b} \pm 0.012_{\xi_\perp(0)} \pm 0.0057_{\xi_\parallel(0)} \pm 0.083_{\rm rest} )  \label{ratioSM1}\\
&=& -0.701 \pm 0.15\; [21\%]\; . \label{ratioSM}
\eea
where the suffix ``rest'' stands for variations of $m_b$, $m_c$, $f_B$, $f_{K^*}^{\perp,\parallel}$, $a_{1,2,}^{\perp,\parallel}$ and $\lambda_B$. Note, in particular, that form factors uncertainties on ${\cal R}[q_0^2]$ are vanishingly small while the impact of all other sources of uncertainty is enhanced. The basic reason behind this behavior is that the effect of any parameter with the exception of the two soft form factors is a parallel shift of the whole forward--backward asymmetry spectrum, thus enhancing their impact on the ratio ${\cal R}[q_0^2]$. Adopting the narrower variation $\mu_b \in [4,5.6]\; {\rm GeV}$ suggested in Ref.~\cite{Altmannshofer:2008dz}, the scale uncertainties in Eqs.~(\ref{lowerSM1}, \ref{higherSM1}, \ref{ratioSM1}) shift from $(8.5\%,8.9\%,19\%)$ to $(2\%,2\%,4\%)$; consequently the total uncertainties in Eqs.~(\ref{lowerSM}, \ref{higherSM}, \ref{ratioSM}) shift from $(0.020,0.015,0.15)$ to $(0.018,0.014,0.07)$. The impact of this more restricted scale variation is fairly small on the integrated FBA in the two bins because the total uncertainty is dominated by the form factors. On the other hand, $\delta {\cal R}[4]$ is controlled by the scale variation and, as we explained above, the scale uncertainties in Eqs.~(\ref{lowerSM1}) and (\ref{higherSM1}) are anticorrelated. There are three main arguments in favor of adopting the observable that we propose.
\begin{figure}
\begin{center}
\includegraphics[width=0.47 \linewidth]{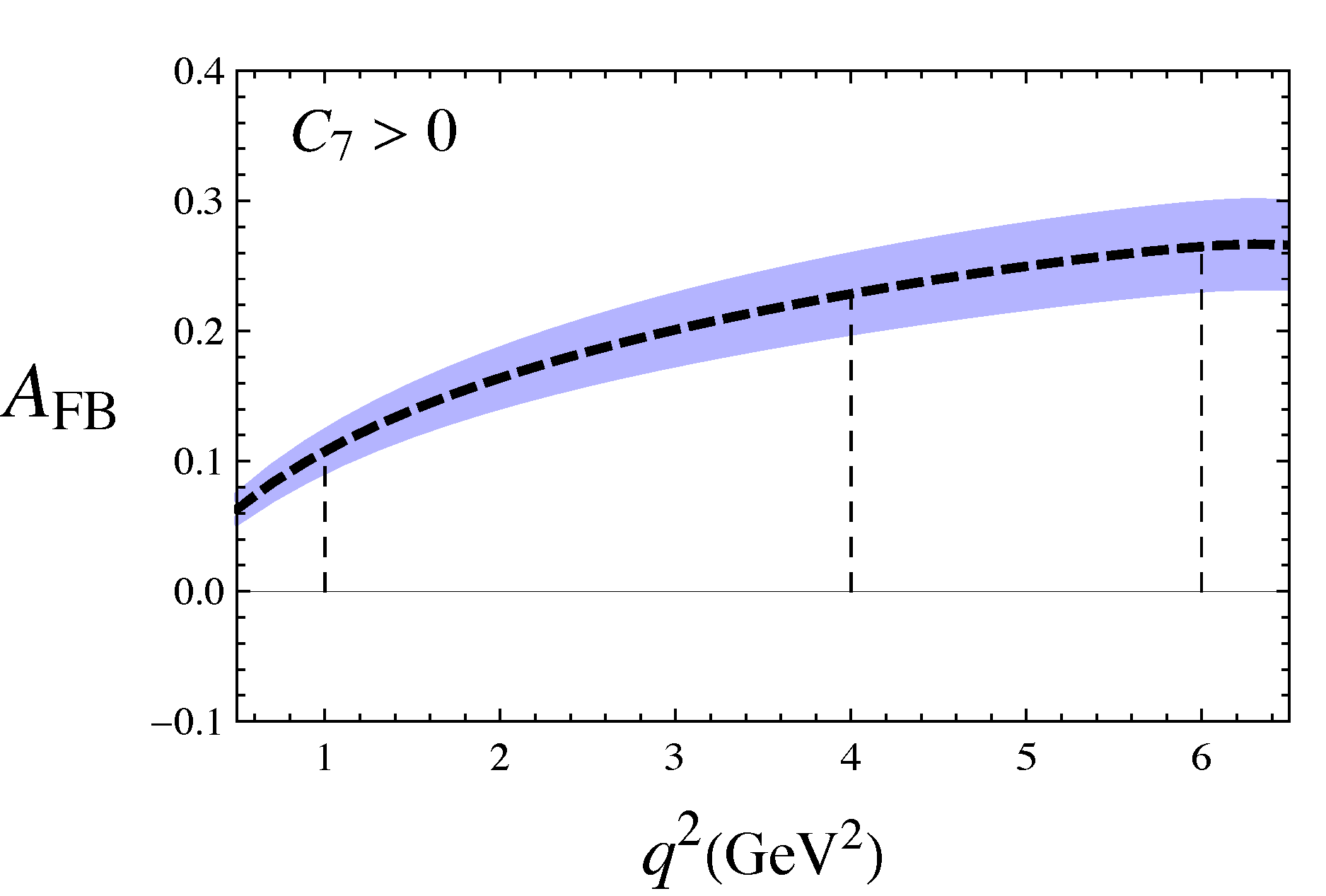}
\end{center}
\vskip -0.7cm
\caption{Error analysis of the forward--backward asymmetry spectrum for $C_7 >0$. \label{fig:AFBpos}}
\end{figure}
\begin{list}{\labelitemi}{\leftmargin=1em}
\item ${\cal R}[q_0^2]$ is sensitive to the same combinations of Wilson coefficients that determine the position of the zero of the spectrum but it is much simpler to determine experimentally. Moreover, it can be measured even in presence of new physics that removes entirely the zero from the spectrum. 
\item The dominant source of uncertainty on ${\cal R}[q_0^2]$ is the scale at which the Wilson coefficients are evaluated. Within the QCD factorization approach this dependence can be reduced by NNLO perturbative calculations and can, therefore, be brought under control. 
\item The branching ratio and the location of zero of the FBA spectrum are completely insensitive to the phase of the Wilson coefficient $C_{10}$. In fact, the FBA is proportional to $C_{10}$ and the location of the zero depends only on $|C_{10}|$. The integrated forward-backward asymmetry dependence on this phase is, on the other end, non trivial. This can be understood by looking once more at the LO expression for the numerator of the forward--backward asymmetry for arbitrary complex Wilson coefficients:
\bea
{\cal A}_{\rm FB} (q^2)  &\propto&   {\rm Re}  \left\{ C_{10}^*  \left( C_9 + Y(q^2) + 2 \f{m_{b} m_B}{q^2} C_7\right) \right\} \\ 
& = & {\rm Re}  \left\{ |C_{10}| \; e^{-i \phi_{10}} \left( |C_9| \; e^{i \phi_9} + Y(q^2) + 2 \f{m_{b} m_B}{q^2} |C_7|\; e^{i \phi_7}\right) \right\} \\ 
&=& |C_{10}| \Big[ \cos (\phi_{10}-\phi_9)\; |C_9| + \cos\phi_{10} \;  {\rm Re} \; Y(q^2) + \cos (\phi_{10} - \phi_7) \; 2 \f{m_{b} m_B}{q^2} |C_7| \nonumber\\
&&+ \sin \phi_{10} \; {\rm Im}\; Y(q^2) \Big]  , \label{numfba}
\eea
where we remind the reader that in the SM we have $\phi_{10} = \phi_9 = 0$ and $\phi_7 = \pi$. Eq.~(\ref{numfba}) displays explicitly the dependence of the integrated FBA on $\phi_{10}$. Unfortunately an effect here can always be reabsorbed into a change of the absolute value and sign of this coefficient (note that the FBA is actually proportional to $|C_{10}|$). We note also that the FBA is a parity odd observable that can be decomposed into CP even and odd components, the latter being the term proportional to $\sin \phi_{10}$. From the inspection of Eq.~(\ref{numfba}) we immediately see that the ratio ${\cal R}[q_0^2]$ is completely insensitive to $|C_{10}|$ (there is a mild residual dependence of ${\cal R}[q_0^2]$ on $|C_{10}|$ through the denominator of the FB asymmetry); in contrast, the dependence on the phase of $C_{10}$ is quite strong and it arises through the interference with the strong phase in the matrix elements of the QCD penguin operators (the imaginary part of $Y(q^2)$ for $q^2 < 4 m_c^2$ originates from the QCD penguin matrix elements involving a light quark loop and is independent of $q^2$ in the limit of vanishing light quark masses~\cite{Bander:1979px}). 

Let us consider the simpler scenario of vanishing new physics weak phases in $C_7$ and $C_9$. In this case we have:
\bea
{\cal A}_{\rm FB} (q^2)  &\propto& |C_{10}| \cos\phi_{10}\Big[ |C_9| +  {\rm Re} \; Y(q^2) + 2 \f{m_{b} m_B}{q^2} |C_7|+ \tan \phi_{10} \; {\rm Im}\; Y(q^2) \Big]  \; . \label{numfba1}
\eea
Integrating over the $[1,4] \; \gev^2$ and $[4,6] \; \gev^2$ ranges and assuming the SM values for the coefficients $C_{7,9}$ we obtain (formulas valid at NLO will be presented in the next section):
\bea
{\cal R}[4] \propto \f{\int_4^6 {\cal A}_{\rm FB} (q^2) \; {\rm d} q^2 }{\int_1^4 {\cal A}_{\rm FB}(q^2) \; {\rm d} q^2 } & \propto & 
- \f{1 +  0.13 \tan \phi_{10} }{1- 0.03 \tan \phi_{10}} \; .
\label{mechanism}
\eea
A very interesting feature of Eq.~(\ref{mechanism}) is that, in this scenario, the difference between ${\cal R}[4]$ and its SM expectation is proportional to $\tan \phi_{10}$. This behavior is usually displayed only by CP violating quantities; for instance, the CP asymmetry in the forward--backward asymmetry studied in Ref.~\cite{Kruger:2000zg} is proportional to ${\rm Im} \; C_{10} = |C_{10}| \sin \phi_{10}$ (see Eq.~(6.10) in Ref.~\cite{Kruger:2000zg}). The ratio ${\cal R}[4]$ retains this dependence on $\phi_{10}$ but is independent of $|C_{10}|$. Therefore we conclude that ${\cal R}[4]$ not only complements the branching ratio and the location of the zero of the spectrum but also provide a unique and direct access to the phase of $C_{10}$. 

\end{list}
\begin{table}
\begin{center}
\begin{tabular}{|c|cccc|} \hline
 & HN & HD& LN & LD \\ \hline
 $q_0$ &   2.672 & 60.55 & -5.425 & 42.97 \\
  $q_{77}$ & 0 & 124.6 & 0 & 38.95 \\
 $q_{88}$ &  0 & 0.7053 & 0 & 0.2085 \\
  $q_{99}$ &  0 & 1.967 & 0 & 1.548 \\
 $q_{1010}$ & 0 & 1.954 & 0 & 1.540 \\
$q_{7}$ & 23.98 & -21.65+8.820 \im & 35.42 & 23.23+3.561 \im \\
  $q_{8}$ &  1.159+1.260 \im & -1.120-0.1590 \im & 1.798+1.859 \im & 1.142+1.521 \im \\
 $q_{9}$ &  2.694 & 9.969+0.3029 \im & 1.993 & 9.110+0.3954 \im \\
$q_{10}$ & -0.6372-0.2986 \im & -16.38 & 1.294-0.3303 \im & -12.92 \\
 $q_{78}$ & 0 & 13.02-13.37 \im & 0 & 3.878-4.161 \im \\
  $q_{79}$ &0 & 17.90 & 0 & 12.03 \\
 $q_{710}$ &-5.718 & 0 & -8.446 & 0 \\
$q_{89}$ &  0 & 1.042+1.088 \im & 0 & 0.6379+0.6671 \im \\
$q_{810}$ &  -0.2763-0.3005 \im & 0 & -0.4288-0.4432 \im & 0 \\
$q_{910}$ & -0.6424 & 0 & -0.4754 & 0 \\ \hline
\end{tabular}
\end{center}
\caption{Coefficients $q_i$ and $q_{ij}$. \label{tab:qij}} 
\end{table}
Let us now discuss scenarios in which the spectrum of the forward--backward asymmetry does not contain a zero. In this case all sources of uncertainties simply shift the spectrum (see Fig.~\ref{fig:AFBpos}) and the ratio ${\cal R}[4]$ turns out to be extremely clean. 
For illustration, we choose $\delta C_7 (\mu_b) \simeq 1.15$ (this value reproduces the central value of the experimental determination of $B\to X_s\gamma$) and obtain:
\bea
\left({\cal R}[4]\right)_{C_7 >0} & = & 1.42 \pm 0.04 \; , \label{ratioC7pos}
\eea
showing significant reduction in the error.  

Finally we summarize the present experimental determination of the $B\to K^* \ell^+\ell^-$ forward--backward asymmetry and attempt a first comparison with the SM prediction. The main limitation we encounter is that the Belle~\cite{:2009zv} and CDF~\cite{cdf10047} collaborations\footnote{The BaBar collaboration published  a study of the integrated FBA in the whole low $q^2$ region~\cite{:2008ju}, without giving separate results for the $q^2$ bins we are interested in in this work.} present results in the whole low dilepton invariant mass region ($0 < m_{\ell \ell} < 8.68 \; \gev^2$) including regions (i.e. $m_{\ell\ell} < 1 \; \gev^2$ and $m_{\ell\ell} > 6 \; \gev^2$) in which the theory approach we are adopting is subject to larger uncertainties. The Belle and CDF results, their weighted average and the corresponding SM predictions, with the caveat we just mentioned, read (the suffixes refer to the $q^2$ bin expressed in $\gev^2$):
\begin{center}
\begin{tabular}{|c|ccc|} \hline 
\vphantom{$\Big($} & ${\cal A}_{\rm FB}^{[0,2]}$ & ${\cal A}_{\rm FB}^{[2,4.3]}$ & ${\cal A}_{\rm FB}^{[4.3,8.68]}$ \\[0.2cm] \hline
\vphantom{$\Big($}Belle & $0.47^{+0.26}_{-0.32}\pm 0.03$& $0.11^{+0.31}_{-0.36}\pm 0.07$ & $0.45^{+0.15}_{-0.21}\pm 0.15$\\[0.1cm]
CDF & $0.13^{+1.65}_{-0.75}\pm 0.25$ & $0.19^{+0.40}_{-0.41}\pm 0.14$& $-0.06^{+0.30}_{-0.28}\pm 0.05$\\[0.1cm]
Average & $0.45 \pm 0.28$& $0.12\pm 0.27$& $0.24\pm 0.23$\\[0.1cm] \hline
\vphantom{$\Big($}SM & $-0.027 \pm 0.004$ & $-0.050 \pm 0.012$ & $0.133\pm 0.026$ \\[0.1cm] \hline
\end{tabular}
\end{center}
Given our ignorance of the correlations between the experimental uncertainties in the various bins, we are unable to extract meaningful constraints on the ratios we are interested in.

\section{Model independent analysis}
\label{sec:mia}
Let us start by presenting a numerical formula that allows to calculate the integrated observable $R[4\; \gev^2]$ for arbitrary values of the complex Wilson coefficients $C_{7,8,9,10} (\mu_b) = C_{7,8,9,10}^{\rm SM} (\mu_b)+ \delta C_{7,8,9,10}$. We write:
\bea
R[4] & = & \frac{f_{\rm HN}}{f_{\rm HD}}/  \left( \frac{f_{\rm LN}}{f_{\rm LD}} \right) \\
f_q & = & q_0 +\sum_{i=7}^{10} {\rm Re} [q_i \; \delta C_i] + 
\sum_{i \leq j=7}^{10} {\rm Re}[ q_{ij} \; \delta C_i \; \delta C_j^*] 
\eea
where in the suffixes AB, ${\rm A} = {\rm L, H}$ stands for the low ($[1,4] \gev^2$) or high ($[4,6] \gev^2$) bin and ${\rm B} = {\rm N, D}$ stands for the numerator or denominator of the forward--backward asymmetry; therefore $f_{\rm HN}$ ($f_{\rm LN}$) and $f_{\rm HD}$ ($f_{\rm LD}$) correspond to the integral of the numerator  and denominator of the forward--backward asymmetry in the $[4,6]  \gev^2$ ($[1,4] \gev^2$) range. The complex coefficients $q_i$ and $q_{ij}$ are given in Table~\ref{tab:qij}. 

In the following we will adopt the numerical formulas for ${\rm BR} (B\to X_s \gamma)$ and ${\rm BR} (B\to X_s \ell^+\ell^-)$ presented in Refs.~\cite{Lunghi:2006hc,Huber:2005ig}; the experimental determinations of these two branching ratios can be found in Refs.~\cite{Chen:2001fja,Abe:2001hk,Aubert:2005cua,Aubert:2006gg,Aubert:2007my,Limosani
,Aubert:2004it,Iwasaki:2005sy} and their world averages are given, for instance, in Refs.~\cite{Barberio:2008fa,Artuso:2009jw,Huber:2008ak}.
\begin{figure}
\begin{center}
\begin{tabular}{ll}
\includegraphics[width=0.4 \linewidth]{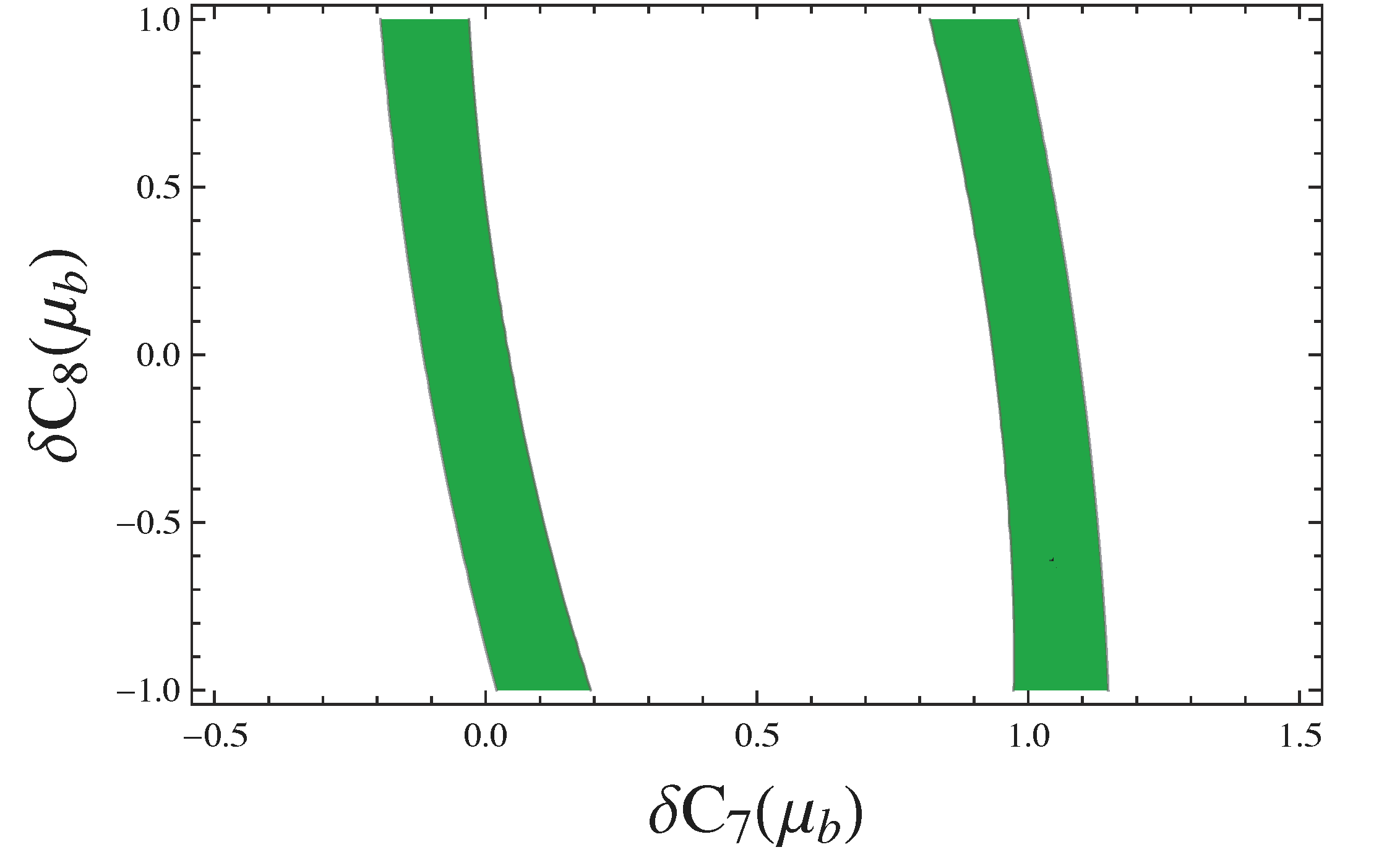}  &
\includegraphics[width=0.4 \linewidth]{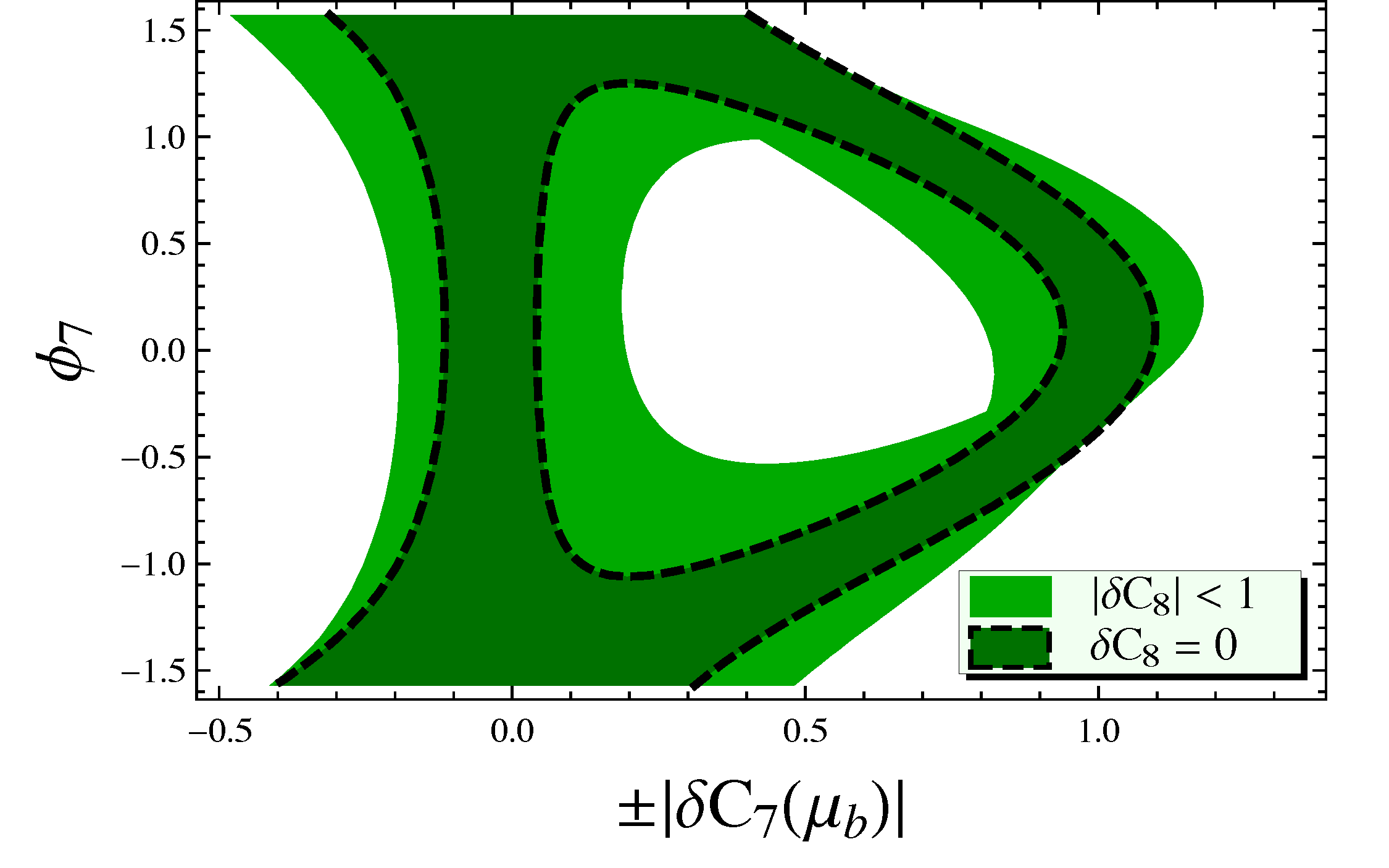}\cr
\includegraphics[width=0.44 \linewidth]{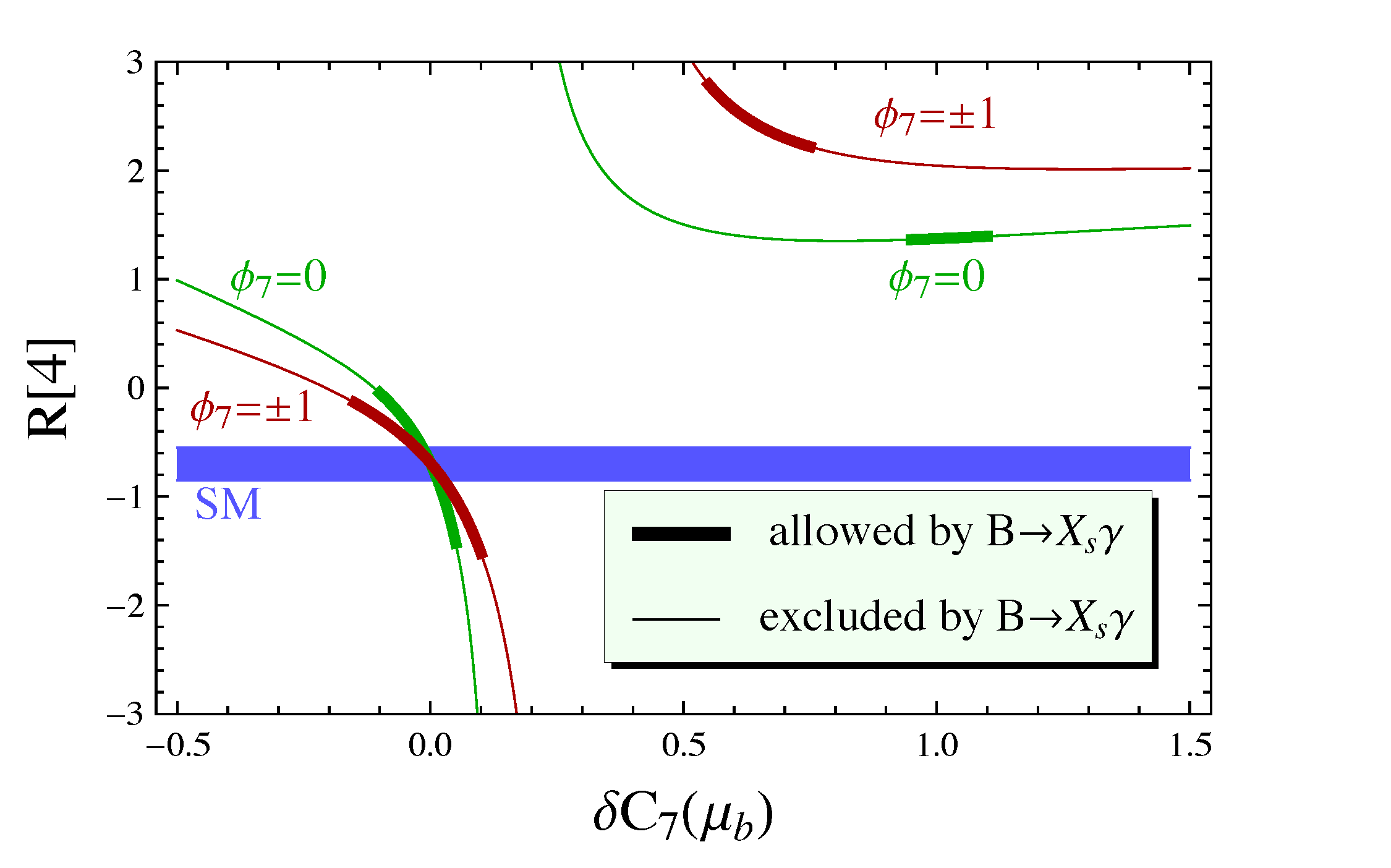} & 
\includegraphics[width=0.44 \linewidth]{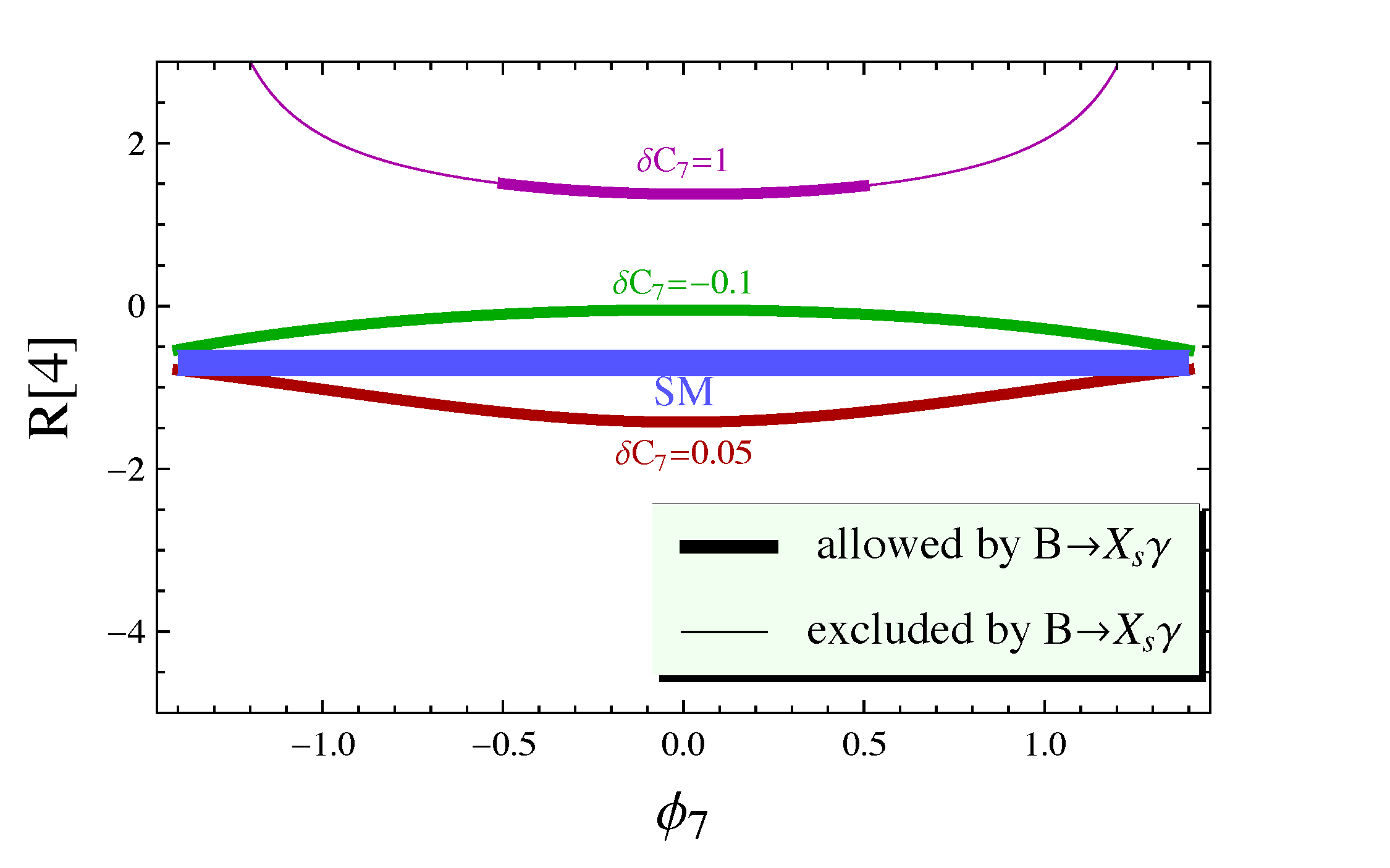} \cr
\end{tabular}
\end{center}
\vskip -0.7cm
\caption{{\it Upper panels}. 95\% C.L. regions allowed by $B\to X_s \gamma$ in the $[\delta C_7,\delta C_8]$ (for real $\delta C_i$) and complex $\delta C_7$ planes. {\it Lower panels}. Dependence of ${\cal R}[4]$ on the Wilson coefficients $C_{7}$. \label{fig:c7}}
\end{figure}

\begin{figure}
\begin{center}
\includegraphics[width=0.32 \linewidth]{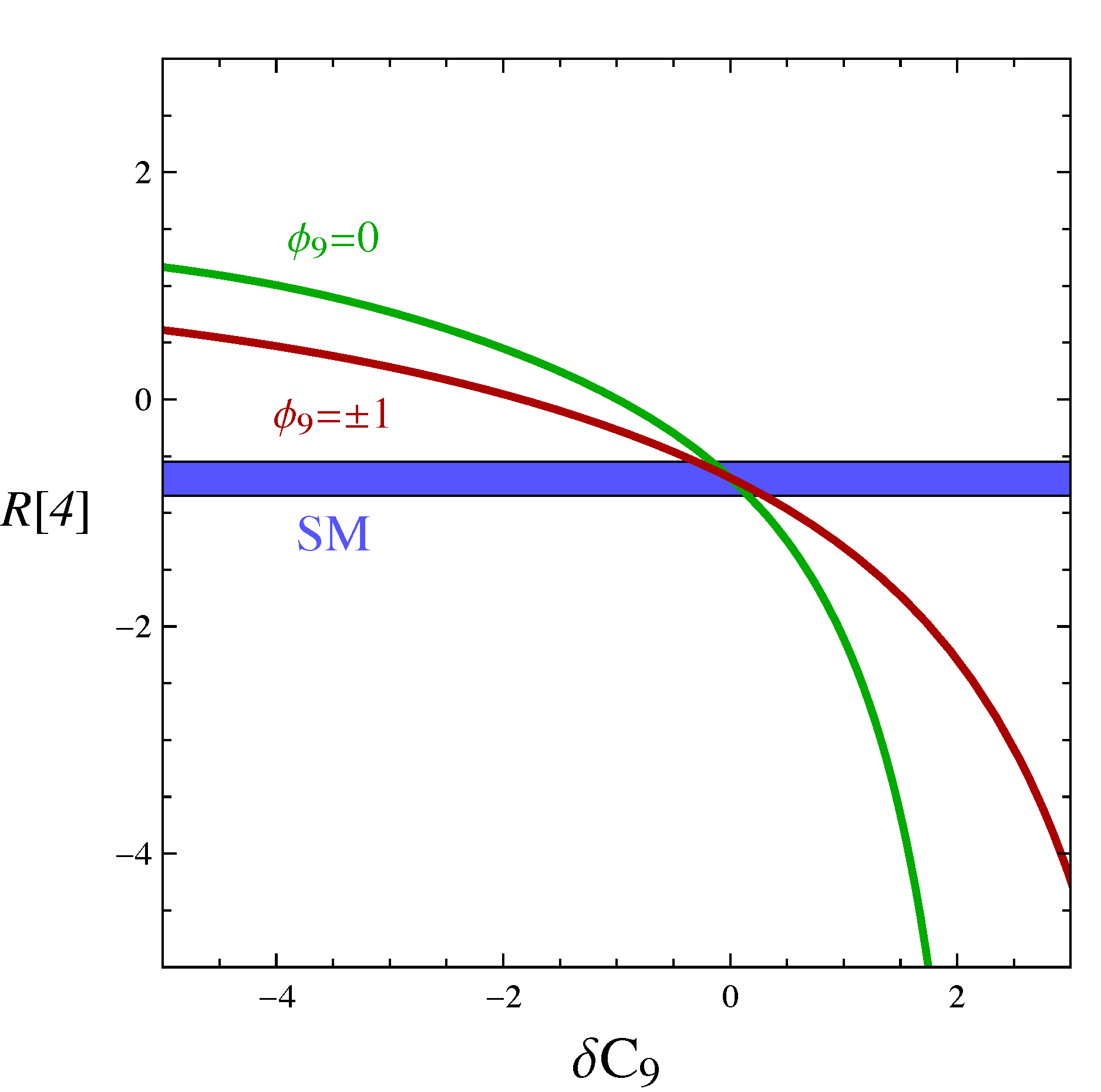}
\includegraphics[width=0.32 \linewidth]{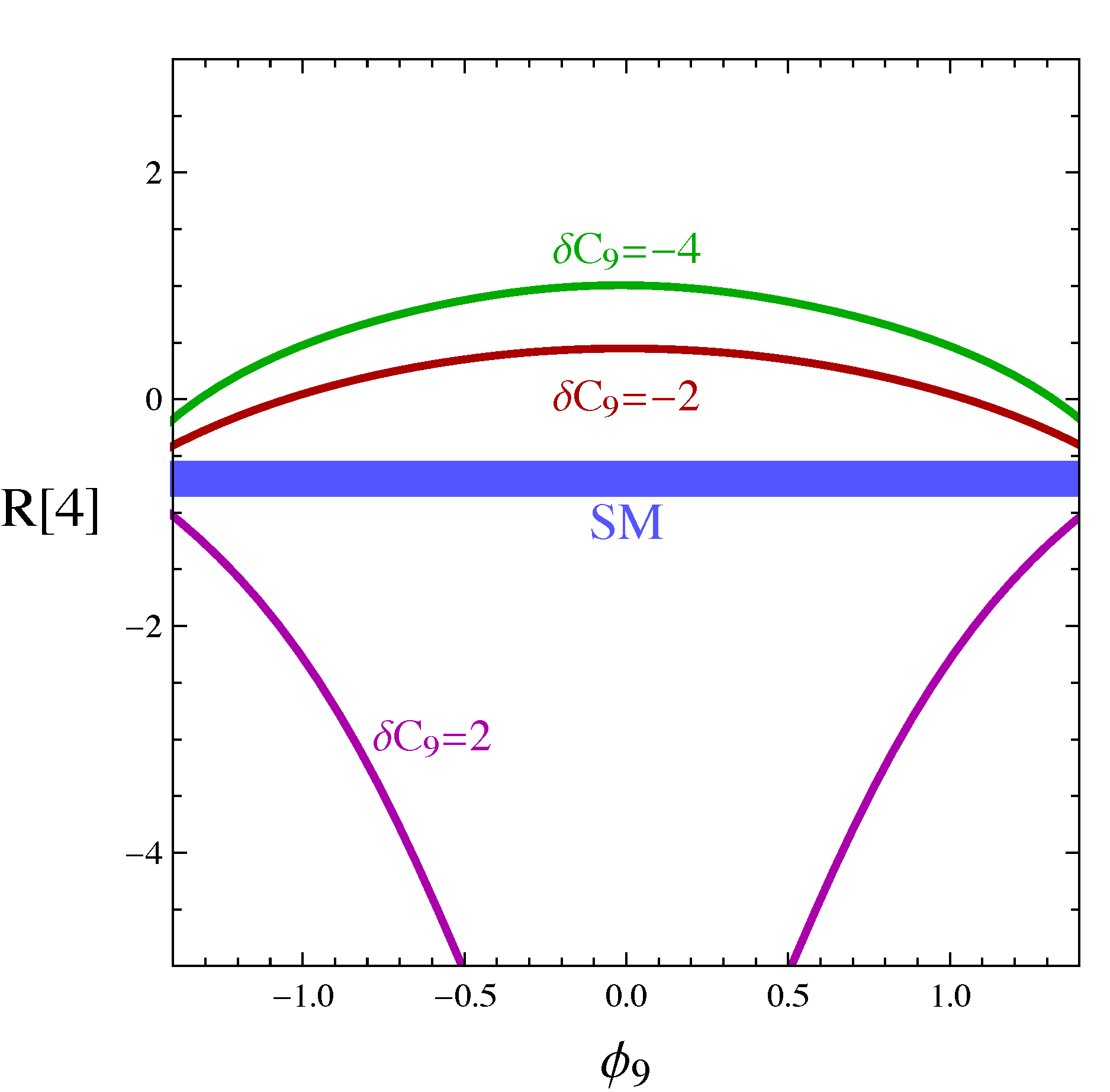}
\includegraphics[width=0.32 \linewidth]{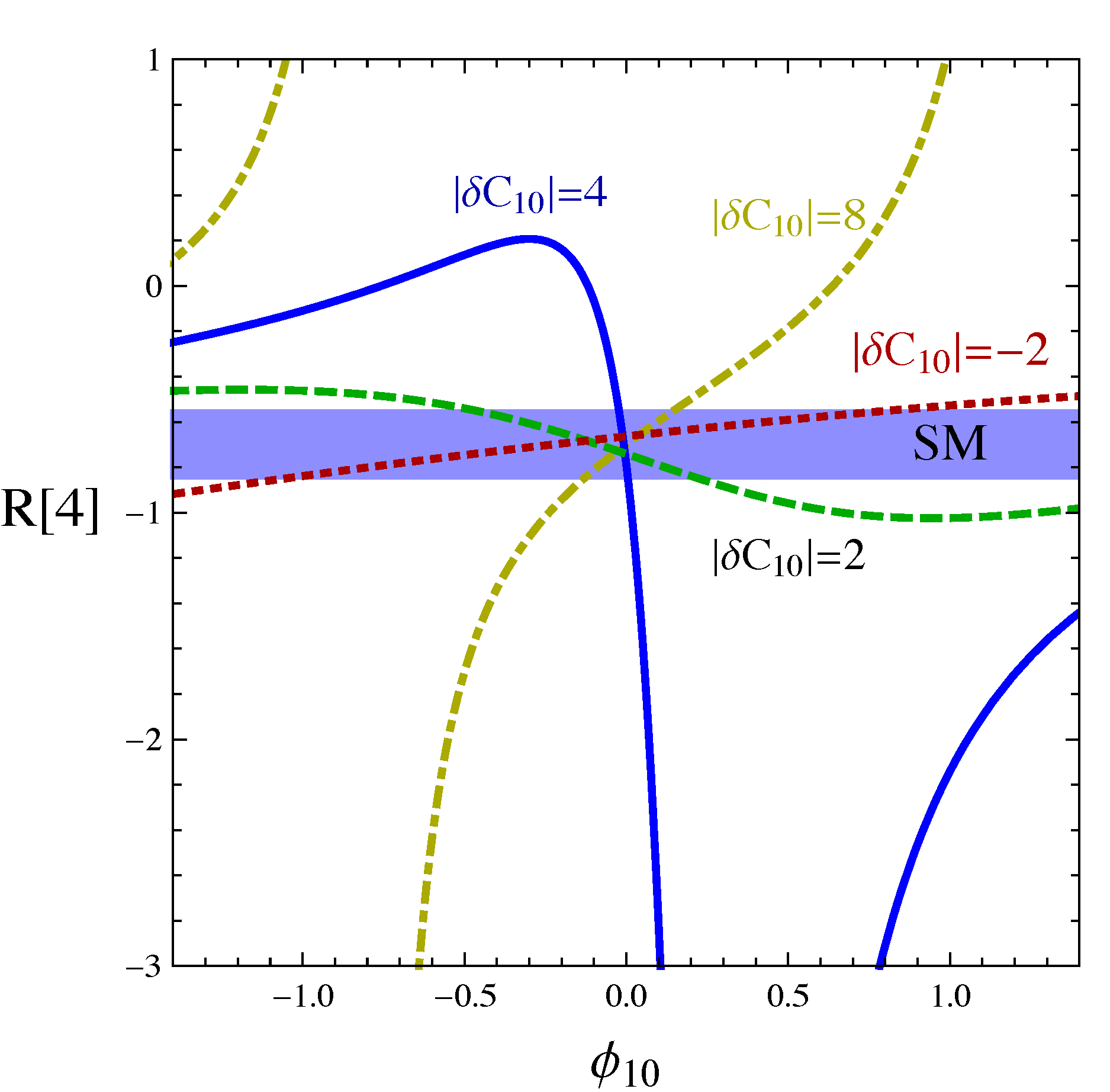}
\includegraphics[width=0.35 \linewidth]{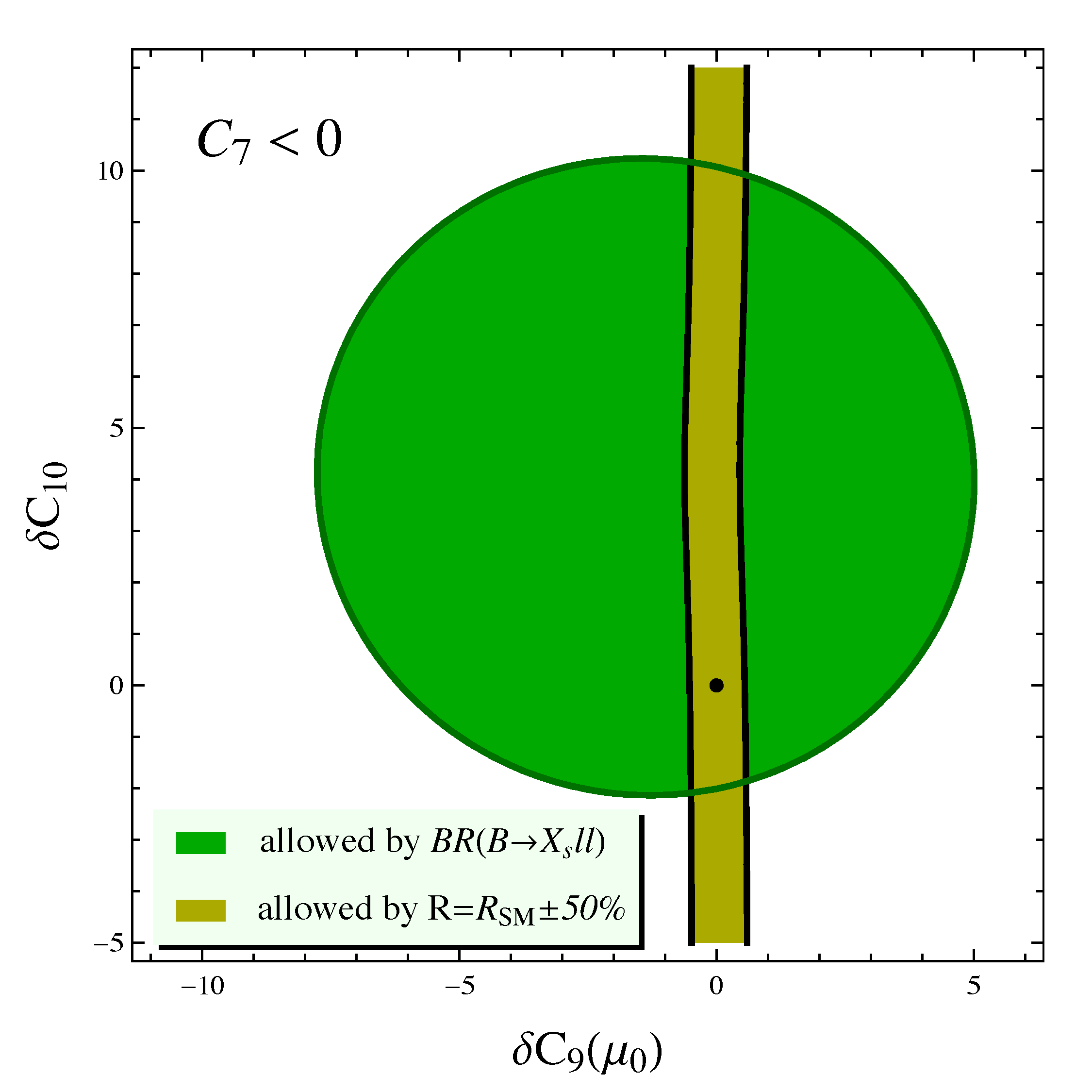}
\includegraphics[width=0.35 \linewidth]{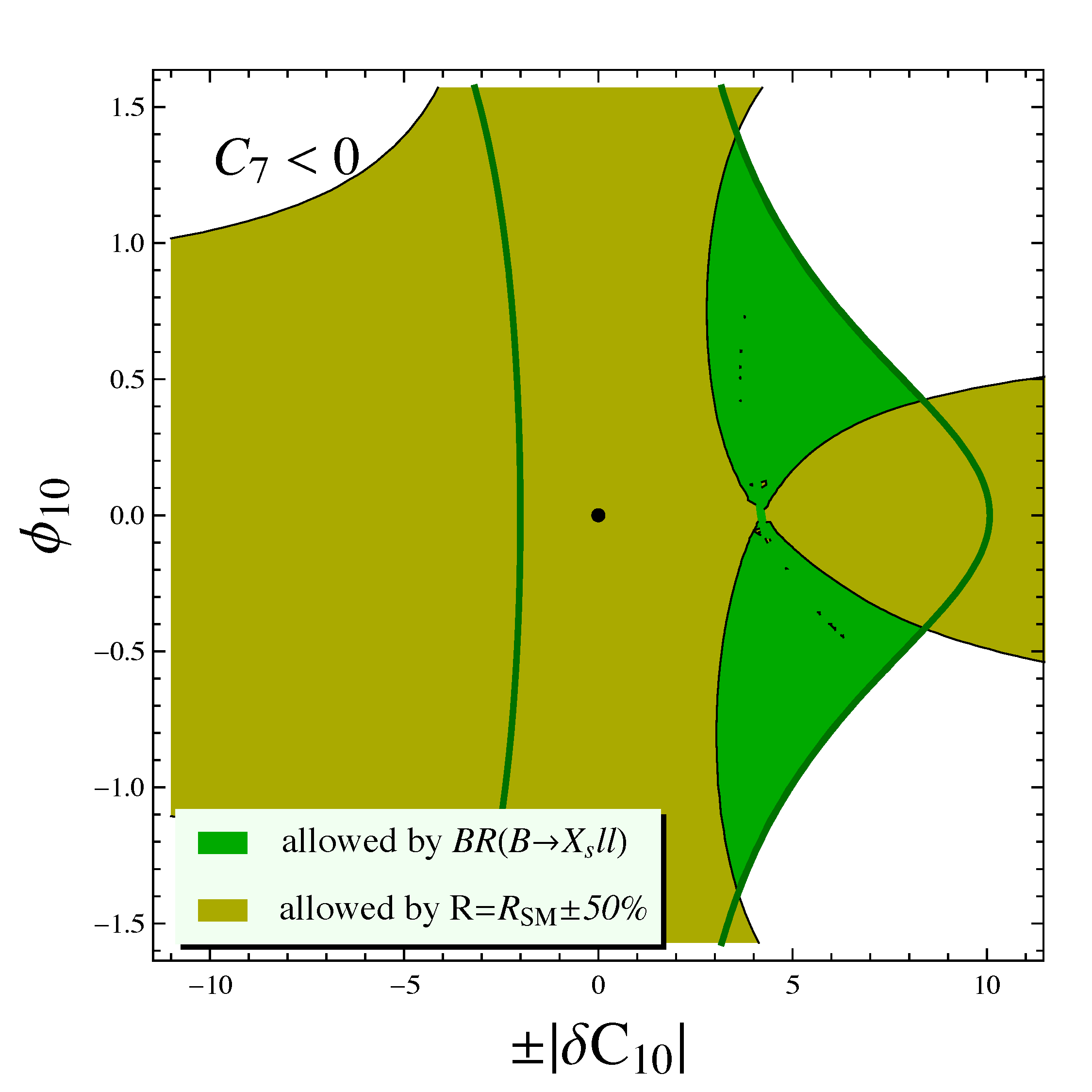}
\end{center}
\vskip -0.7cm
\caption{{\it Upper panels}. Dependence of ${\cal R}[4]$ on the Wilson coefficients $C_9$ and $C_{10}$. {\it Lower panels}. Bounds on the Wilson coefficients $C_{9,10} (\mu_b)$ from $B\to X_s \ell\ell$ (dark green regions) and from a possible measurement of the ratio $R[4]$ (light green regions). We fix $C_7$ to its SM value and assume $R[4]_{\rm exp} = R[4]_{\rm SM} \pm 50\%$.
\label{fig:c9c10}}
\end{figure}

As we discussed in the previous section, although ${\cal R}[q_0^2]$ is an intrinsically CP conserving observable, the interference between weak and strong phases in the numerator of the forward--backward asymmetry introduces a strong sensitivity to new physics CP violating phases\footnote{The impact of CP violating phases in the denominator of the FBA, i.e. the branching ratio, is much smaller and essentially confined to the interference between $C_7$ and $C_9$.}. Especially interesting is the sensitivity to the elusive phase of $C_{10}$. We present the result of a model independent study of new physics contributions to $C_7$, $C_9$ and $C_{10}$ in Figs.~\ref{fig:c7}-\ref{fig:c9c10}. When allowing for complex contributions to $\delta C_i$, we force the phases to lie in the $[-\pi/2,\pi/2]$ range and write $\delta C_i = \pm |\delta C_i| e^{i \phi_i}$ where $\phi_i = \arg \delta C_i$ or $(\arg \delta C_i -\pi)_{{\rm \; mod}\;  2\pi}$. 
In the upper panels of Fig.~\ref{fig:c7} we show bounds implied by $B\to X_s \gamma$ on the NP contributions $\delta C_7 (\mu_b)$ and $\delta C_8 (\mu_b)$. In the left panel we consider real coefficients and in the right one we entertain the complex $\delta C_7$ scenario. The two distinct regions that are allowed in the left panel correspond to the two possible signs for the $b\to s \gamma$ amplitude that is approximatively proportional to $C_7 (\mu_b) = C_7^{\rm SM} (\mu_b) + \delta C_7 (\mu_b)$, with $C_7^{\rm SM} \mu_b)<0$. The precise experimental determination of ${\rm BR} (B\to X_s \gamma)$ implies that $\delta C_7$ can lie either in a small region around 0 (same sign solution, $C_7 \sim C_7^{\rm SM} < 0$) or in a narrow band around $\delta C_7 \sim 1$ (opposite sign solution, $C_7 \sim - C_7^{\rm SM} > 0$). In the two lower panels of Fig.~\ref{fig:c7} we show the dependence of ${\cal R}[4]$ on the complex coefficient $\delta C_7$; the parts of the curves that are actually allowed by the $B\to X_s \gamma$ constraint are indicated by thicker lines. From the inspection of these plots we see that these effects do not depend much on possible NP phases in $\delta C_7$ and that for the same sign solution $C_7 <0$, moderately large effects are possible. The opposite sign solution $C_7>0$ yields a change in the sign of ${\cal R}[4]$ but is currently disfavored by $B\to X_s \ell^+ \ell^-$ data.

In Fig.~\ref{fig:c9c10} we present a similar analysis for NP effects on $C_9$ and $C_{10}$. Note that, in the absence of extra phases, ${\cal R}[4]$ is almost completely insensitive to $|\delta C_{10}|$. Note the contrast between complex effects in $C_9$ and $C_{10}$. In the former case the impact of a non-vanishing $\phi_9$ can be sizable but not essential in order to obtain large effects on ${\cal R}[4]$; in the latter case, the condition $\phi_{10} \neq 0$ is necessary in order to have an effect at all.

An interesting feature of these plots is the appearance of very large effects on ${\cal R}[4]$ for certain values of the coefficients $C_{7,9}$. Positive new physics contributions to $C_{7,9}$ tend to shift the zero of the FB asymmetry towards smaller $q^2$ values. The integral of ${\cal A}{\rm FB} (q^2)$ in the $[1,4]\; \gev^2$ range receives, therefore, positive and negative contributions and, for some value of the Wilson coefficients, it can vanish. At this point the ratio ${\cal R}[4]$ can become very large and one may need to adopt a different strategy (e.g. consider ${\cal R}[q^2]$ for some other $q^2$ value).

It is particularly interesting to extract the bounds on $\delta C_9$ and $\delta C_{10}$ that follow from a hypothetical measurement of ${\cal R}[4]$. In the lower panels of Fig.~\ref{fig:c9c10} we assume an experimental determination of the ratio ${\cal R}[4]$ with an uncertainty of $50\%$ and central values agreeing with the SM and study its impact on the real $[\delta C_9, \delta C_{10}]$ and in the complex $\delta C_{10}$ planes. In these plots the shaded disks represent the area allowed by ${\cal B} (B\to X_s \ell\ell)$ and the light shaded areas are the region of the $[C_9,C_{10}]$ plane that is selected by the measurement of ${\cal R}[4]$. 

From the inspection of the lower-left panel of Fig.~\ref{fig:c9c10} we conclude that even a relatively poor determination of ${\cal R}[4]$ can help hugely in determining the value of $C_9$. The lower-right panel of Fig.~\ref{fig:c9c10} illustrates the impact of a complex phase in $\delta C_{10}$: as long as $\phi_{10} \neq 0$ a determination of ${\cal R}[4]$ strongly constraints $|\delta C_{10}|$.  
\section{Few concrete examples}
\label{sec:examples}
In this section we briefly entertain three distinct new physics scenarios and investigate the role of a measurement of ${\cal R}[4]$ on their phenomenology. In Sec.~\ref{sec:4th} we consider the addition of a fourth generation to the SM. In Sec.~\ref{sec:mssm} we discuss one supersymmetric extension of the SM: an $R$-parity conserving MSSM with extra sources of flavor changing interactions in the down squark sector. In Sec.~\ref{sec:zprime} we consider a family--dependent $U(1)^\prime$ model. Note that in the 4th generation and $Z^\prime$ models the effects are driven by new CP violating phases in $C_{10}$; in the SUSY scenario, the effect is controlled by more traditional contributions to the (chromo--)magnetic moment coefficients $C_{7,8}$.

\subsection{4th generation}
\label{sec:4th}
The inclusion of a fourth generation is perhaps the simplest extension of the Standard Model. The phenomenology of this model, usually referred as ${\rm SM}4$, has recently been the subject of intensive scrutiny (see, for instance, Refs.~\cite{Kribs:2007nz,Soni:2008bc,Holdom:2009rf,Hou:2010mm,Bobrowski:2009ng,Eberhardt:2010bm,Chanowitz:2009mz,Eilam:2009hz,BarShalom:2010bh,Soni:2010xh,Buras:2010pi,Buras:2010nd,Buras:2010cp, Choudhury:2010ya,Chanowitz:2010bm}). In this work we are interested in possible 4th generation effects on $B\to K^* \ell\ell$, specifically on the ratio ${\cal R}[4]$. The two parameters that control ${\rm SM}4$ effects on the magnetic and semileptonic coefficients are the mass of the 4th generation top quark, $m_{t^\prime}$, and the ratio of ${\rm CKM}_4$ elements $\lambda_{tt^\prime}^s \equiv V_{t^\prime b}^{} V_{t^\prime s}^{}/V_{tb}^{}V_{ts}^{*}$. From the analyses in Refs.~\cite{Soni:2010xh,Buras:2010pi} we extract ranges for these two parameters that are phenomenologically viable; in the numerical analysis we take $m_{t^\prime} = 500 \; \gev$ and $|\lambda_{tt^\prime}^s|<0.1$.

The ${\rm SM}4$ expressions for the $t^\prime$ contributions to the Wilson coefficients $C_{7-10}$ can be obtained trivially from the corresponding SM expressions. Since we are interested in the forward--backward asymmetry in $B\to K^*\ell\ell$, we do not need separately the combinations $V_{t^\prime b}^{} V_{t^\prime s}^{}$ and $V_{tb}^{}V_{ts}^{*}$, but only their ratio.
\begin{figure}
\begin{center}
\includegraphics[width=0.47 \linewidth]{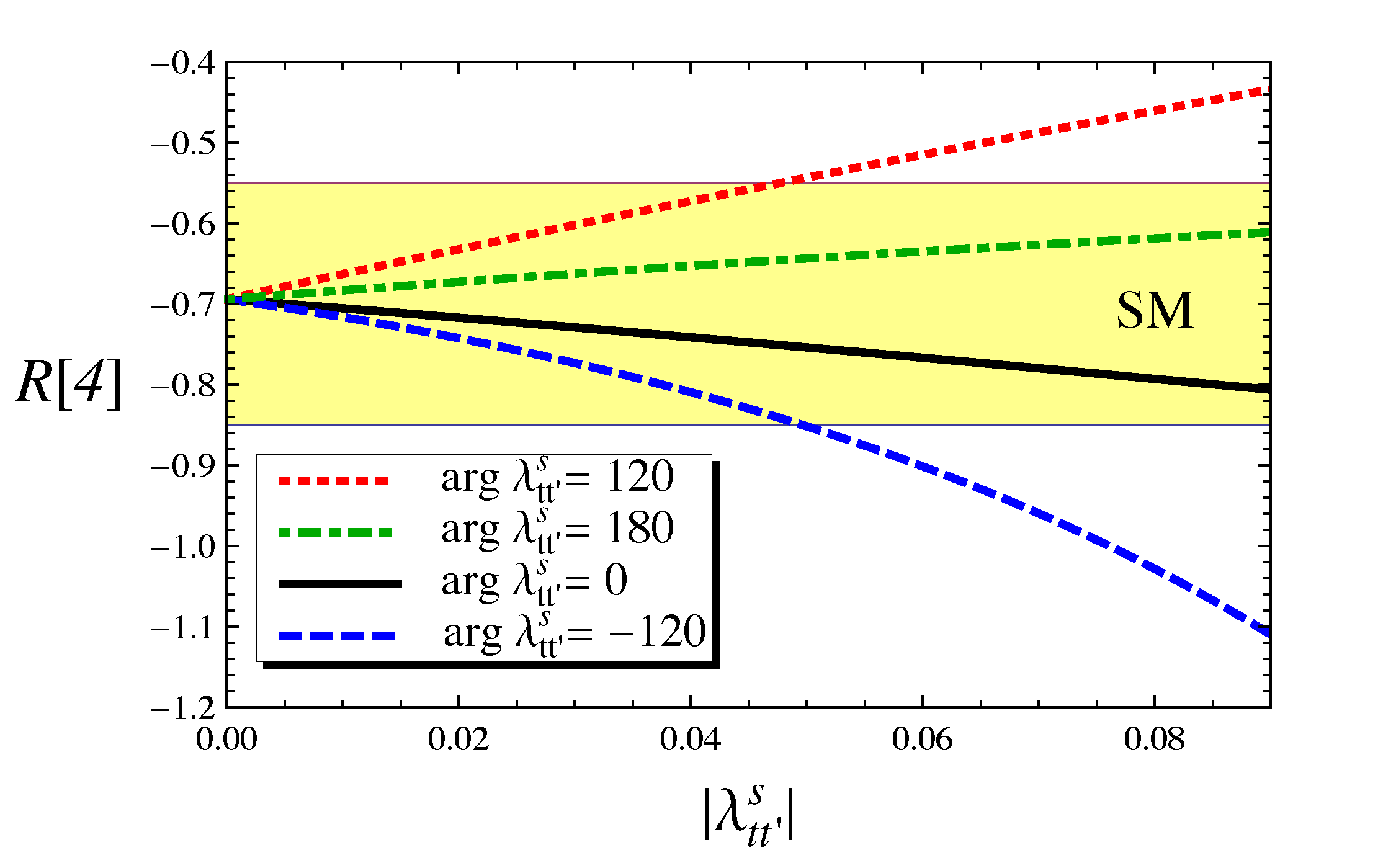}
\includegraphics[width=0.47 \linewidth]{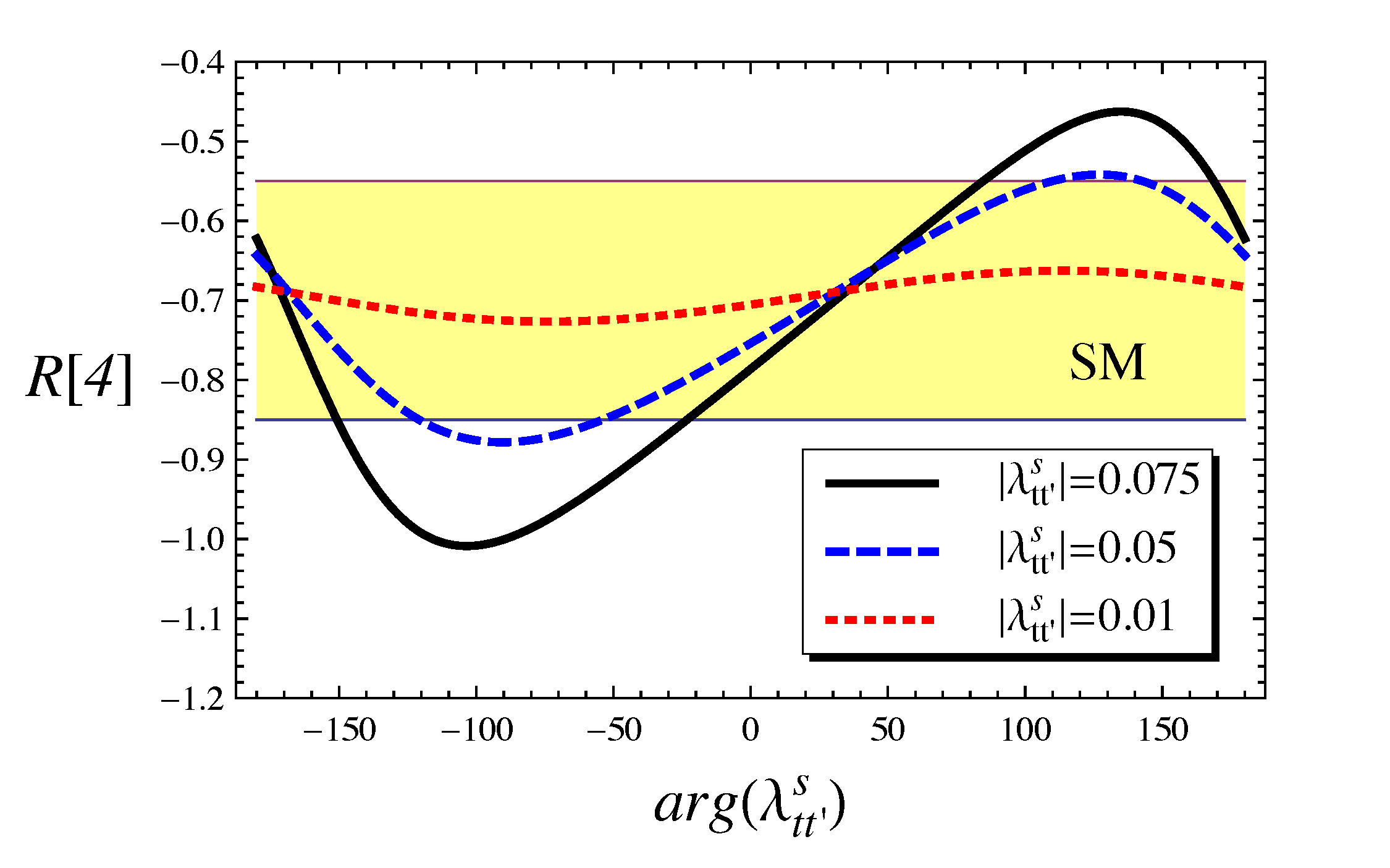}
\end{center}
\vskip -0.7cm
\caption{Expectations for ${\cal R}[4]$ in ${\rm SM}4$. We fix the mass of the $t^\prime$ to $500 \; \gev$ and allow the ratio of ${\rm CKM}_4$ elements $\lambda_{tt^\prime}^s$ to vary according to the bounds obtained in Refs.~\cite{Soni:2010xh,Buras:2010pi}. \label{fig:4th}}
\end{figure}

In Fig.~\ref{fig:4th} we show the dependence of ${\cal R}[4]$ on the magnitude and phase of $\lambda_{tt^\prime}^s$ for $m_{t^\prime} = 500 \; \gev$. It is clear that large deviations are possible though only in presence of large phases in the new ${\rm CKM}_4$ elements. 

This strong dependence on the CP violating phase of $\lambda_{tt^\prime}^s$ stems entirely from the 4th generation contributions to $C_{10}$. In fact, the functional dependence of $\delta C_9^{\rm 4th}$ and $\delta C_{10}^{\rm 4th}$ on $m_{t^\prime}$ is such that for large $t^\prime$ masses (in the range $[200,1000] \gev$) one has $2 \lesssim |\delta C_{10}^{\rm 4th}/\delta C_{9}^{\rm 4th} | \lesssim 10$.
\subsection{MSSM}
\label{sec:mssm}
Let us consider an $R$--parity conserving MSSM with non-vanishing sources of flavor changing neutral currents in the down squark sector (i.e. non Minimal Flavor Violating). As an example, we focus on contributions induced by a non-vanishing value of the $LR$ mixing between the second and third generation. Adopting the formalism of Ref.~\cite{Lunghi:1999uk} we define this mixing in the super--CKM basis (a basis in which the squark mass matrices are subject to the same rotations that diagonalize the regular quark sector). 

In this model, gluino vertices are flavor changing and the dominant contributions to the Wilson coefficients come from 1-loop involving gluino and down squarks. The parameter that we need are the gluino and down squark masses and the mass insertion $(\delta_{23}^d)_{LR}$ (see Ref.~\cite{Lunghi:1999uk} for more details). 
\begin{figure}
\begin{center}
\includegraphics[width=0.47 \linewidth]{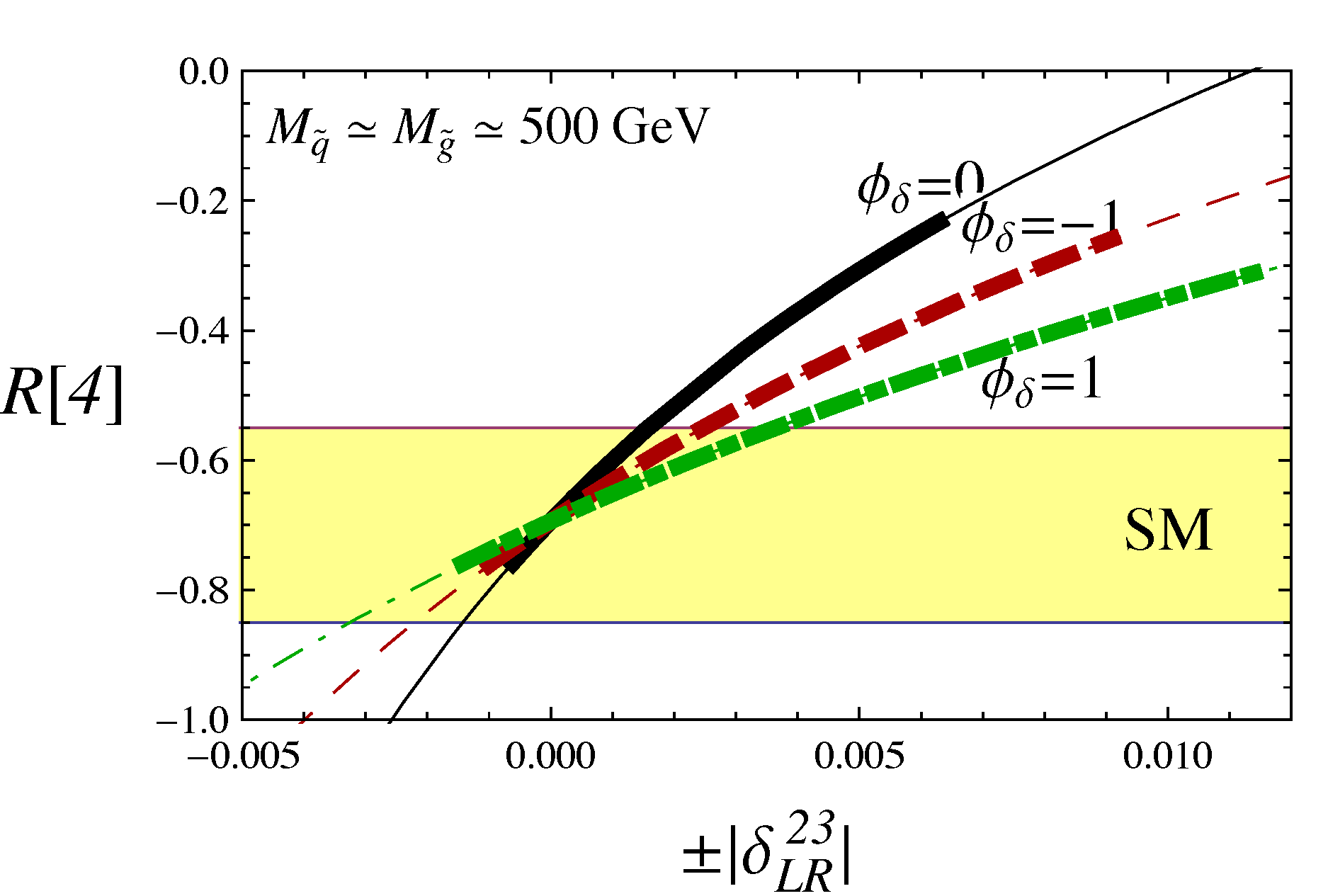}
\includegraphics[width=0.47 \linewidth]{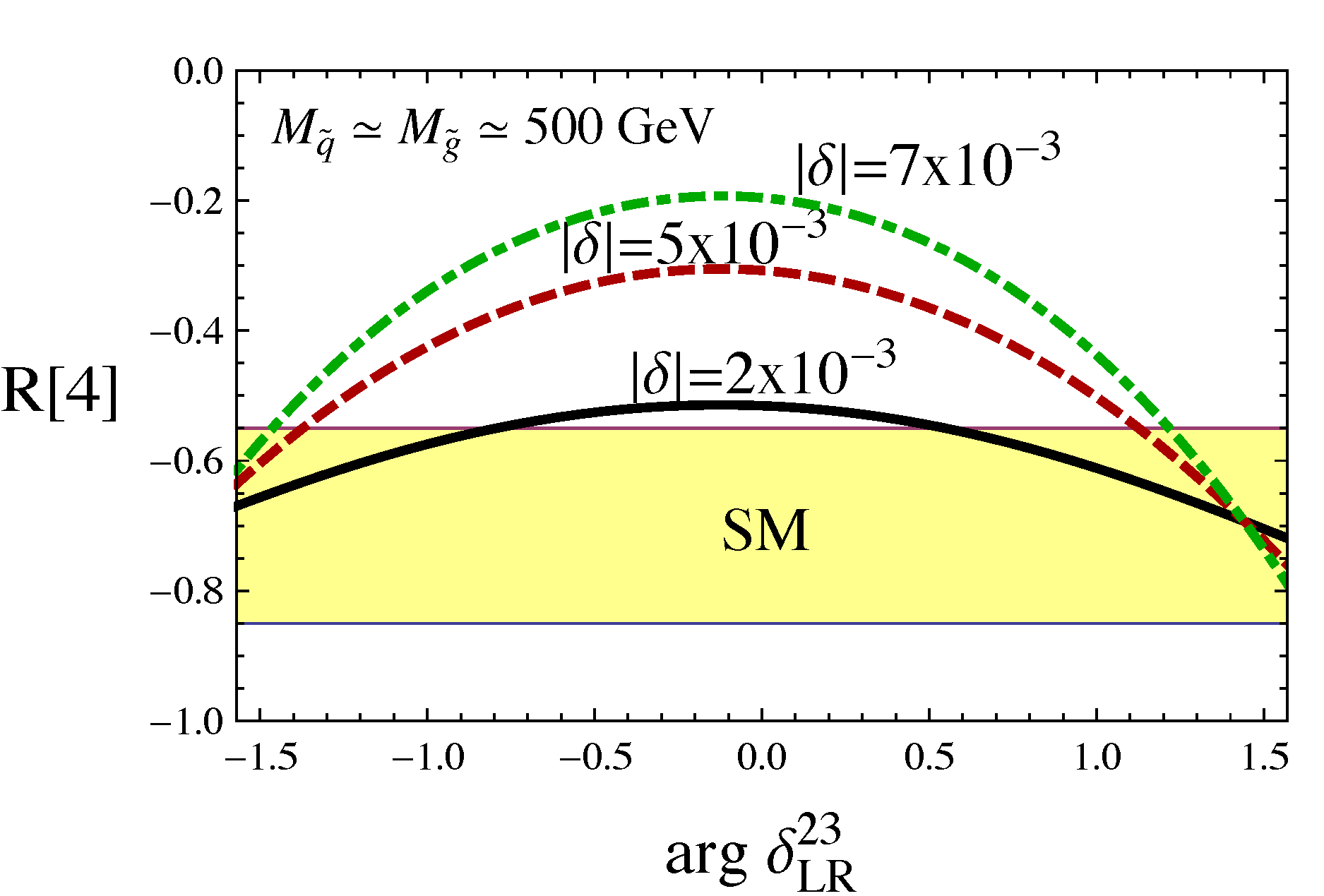}
\end{center}
\vskip -0.7cm
\caption{Dependence of ${\cal R}[4]$ on $|\delta_{23}^{LR}|$ for $m_{\tilde g} \simeq m_{\tilde q} \simeq 500 \; \gev$. We indicate with a thicker line the region allowed by $B\to X_s \gamma$. Solid, dashed and dot-dashed lines correposnd to various values of $\phi_\delta$ and $|\delta_{23}^{LR}|$ . \label{fig:susy}}
\end{figure}

The dominant constraints on this insertion come from $B\to X_s \gamma$, $B\to X_s \ell\ell$, $B_s$ mixing, vacuum stability and absence of color breaking minima in the potential. The interplay of these constraints and the allowed ranges for $(\delta_{23}^d)_{LR}$ have been studied at length in the literature. In this work we use the results of Ref.~\cite{Silvestrini:2007yf}. Note that in Ref.~\cite{Silvestrini:2007yf} the bounds are given for $m_{\tilde q} \simeq m_{\tilde g} \simeq 350 \; \gev$ and can be easily rescaled to cover the $m_{\tilde q} \simeq m_{\tilde g} \simeq 500 \; \gev$ that we consider here.

In Fig.~\ref{fig:susy} we show the size of supersymmetric corrections to ${\cal R}[4]$. The opposite sign solution for the $B\to X_s \gamma$ amplitude (i.e. the $C_7 >0$ scenario) requires a very large mass insertion, $(\delta_{23}^d)_{LR} \sim -0.08 \times (m_{\tilde q}/500 \; \gev)$, that is excluded by the charge and color breaking bound that reads $|(\delta_{23}^d)_{LR}| \lesssim 0.017 \times (m_{\tilde q}/500 \; \gev)$. On the other hand, the dependence of ${\cal R}[4]$ on the phase of the mass insertion is not as strong as in the 4th generation case. The reason for this is that in this supersymmetric model we do not have large contributions to $C_{10}$; therefore, the mechanism (see Eq.~(\ref{mechanism})) that induces a large sensitivity to CP violating phases in ${\cal R}[4]$ is not active in this case. 

\subsection{Family--dependent $U(1)^\prime$}
\label{sec:zprime}
As a final example of new physics scenarios in which the observable ${\cal R}[4]$ has a strong discriminating power, we consider a $Z^\prime$ model with tree--level flavor changing couplings proposed and analyzed in Ref.~\cite{Barger:2009qs} (see also Ref.~\cite{Hiller:2002ci} for an analysis of an effective flavor changing $Z\bar bs$ coupling). In this analysis we assume that the $U(1)^\prime$ coupling constant ($g_2$ in the notation of Ref.~\cite{Barger:2009qs}) is identical to the SM U(1) coupling $g_1$ and that the $Z$ and $Z^\prime$ have identical couplings to leptons. Therefore, the only parameters that remain (for what concerns $b\to s$ transitions) are the $Z^\prime$ mass and the flavor changing complex couplings $B_{bs}^{L,R}$, defined as
\bea
{\cal L}_{FC} & = & - g_2 \;  Z^\prime_\mu \; \bar b \gamma^\mu \left[ 
B_{bs}^L P_L + B_{bs}^R P_R \right] s \; .
\label{zprime-lag}
\eea
\begin{figure}
\begin{center}
\includegraphics[width=0.47 \linewidth]{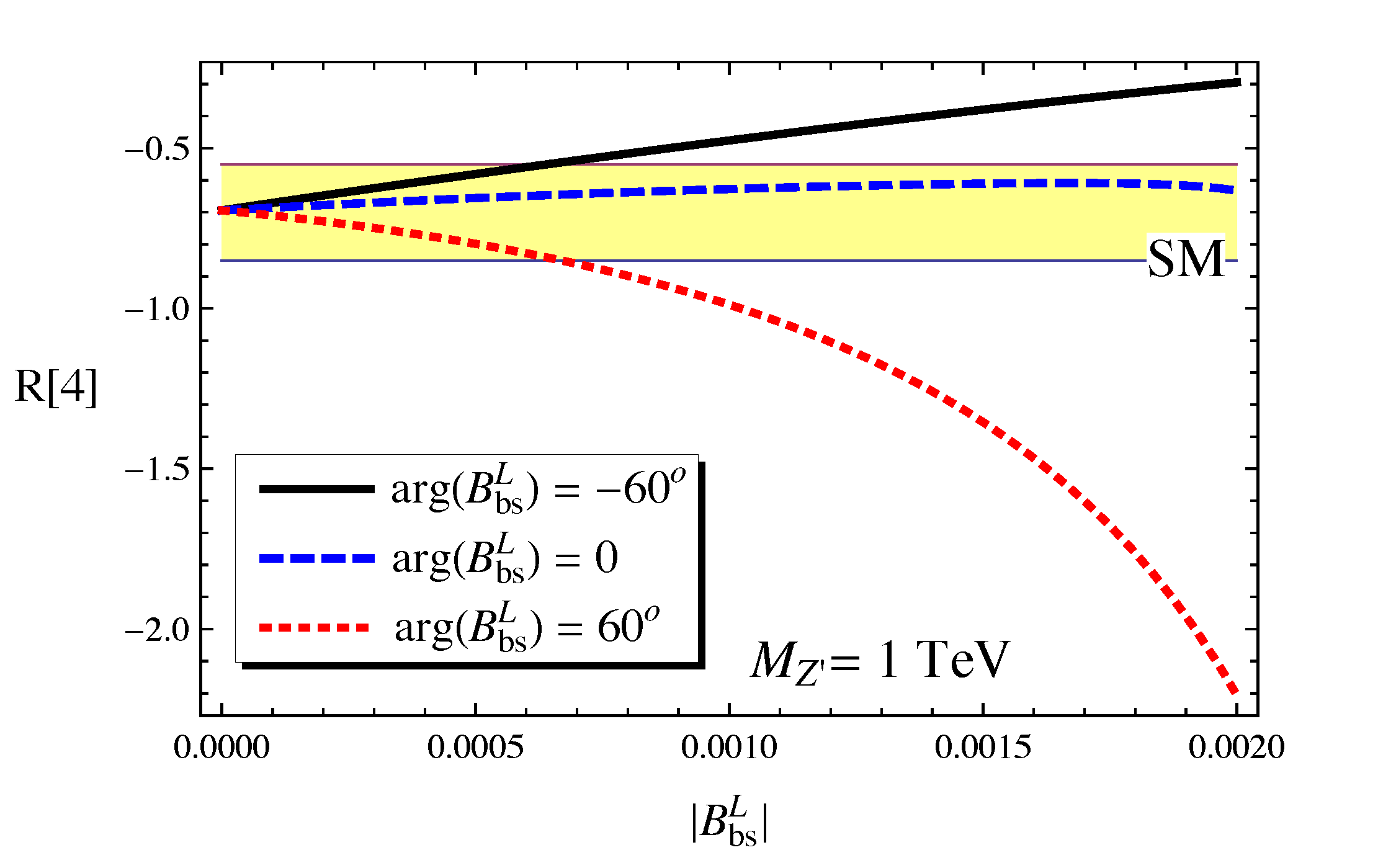}
\includegraphics[width=0.47 \linewidth]{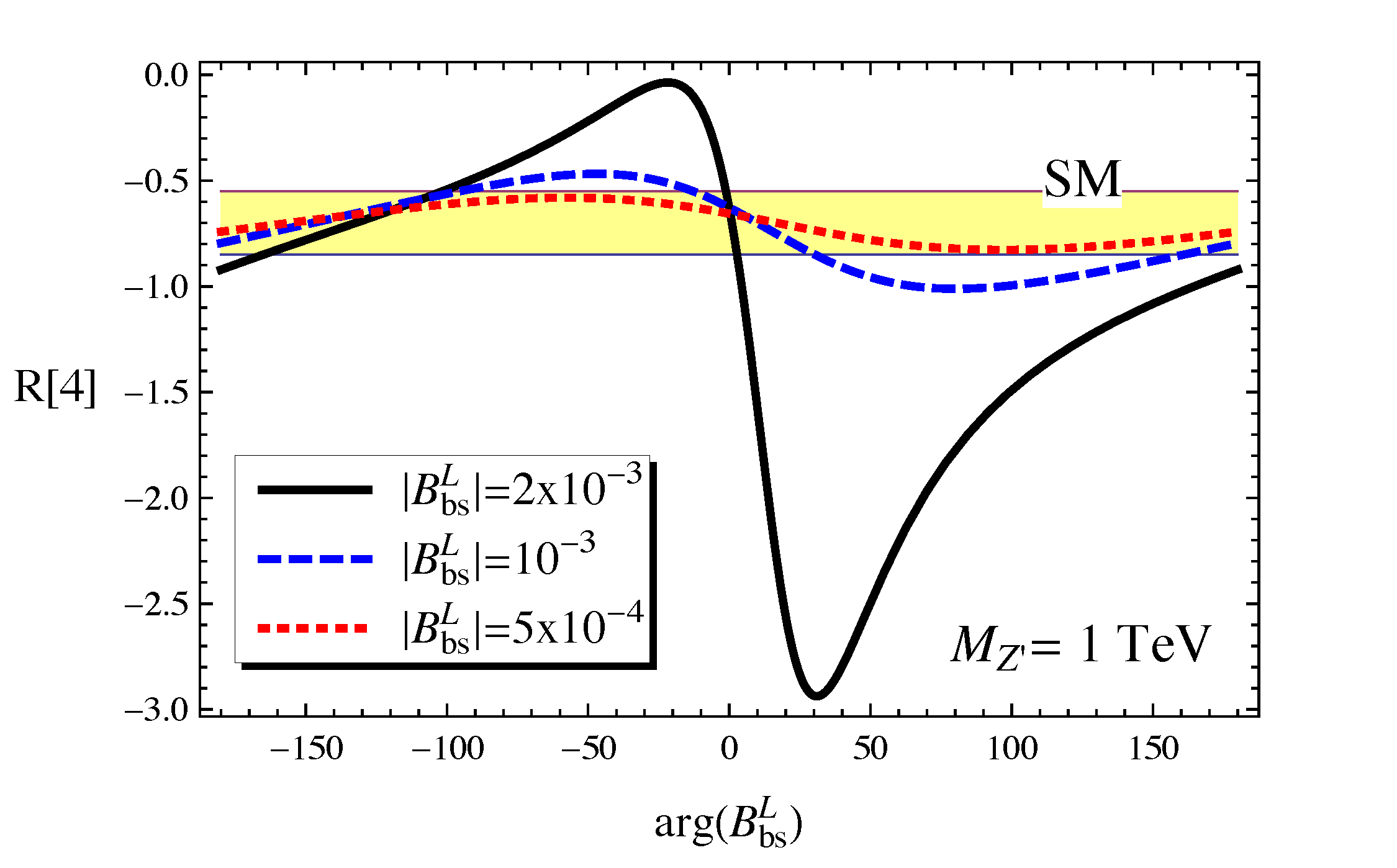}
\end{center}
\vskip -0.7cm
\caption{Expectations for ${\cal R}[4]$ in a family--dependent $U(1)^\prime$ model. We fix the mass of the $Z^\prime$ to $1 \; {\rm TeV}$, allow the complex parameter $|B_{bs}^L| < 2 \times 10^{-3}$~\cite{Barger:2009qs} and assume that the $Z^\prime$ coupling to leptons is identical to the SM. \label{fig:zprime}}
\end{figure}
The effect of $B_{bs}^R$ is to generate contributions to the operators $O_{9,10}^\prime$ obtained from $O_{9,10}$ by replacing the left--handed quark current with a right--handed one. In our study we will not consider this possibility and assume that $Z^\prime$ couplings follow the weak interactions chiral structure (i.e. we take $B_{bs}^R = 0$). Utilizing Eq.~(\ref{zprime-lag}) we obtain:
\bea
\delta C_9 & = & -\frac{ 4 s_W^2 -1}{2} \; \left( V_{tb}^{} V_{ts}^* \frac{\alpha_{\rm em}}{4\pi} \right)^{-1} \frac{g_2 m_Z}{g_1 m_{Z^\prime}}\; \hat B_{bs}^L \; , \label{eq:c9zprime}\\
\delta C_{10} & = & - \frac{1}{2} \;\left( V_{tb}^{} V_{ts}^*\frac{\alpha_{\rm em}}{4\pi}  \right)^{-1} \frac{g_2 m_Z}{g_1 m_{Z^\prime}}\;  \hat B_{bs}^L \; , \label{eq:c10zprime}
\eea
where $s_W$ is the weak mixing angle and, following Ref.~\cite{Barger:2009qs}, we defined\footnote{In Ref.~\cite{Barger:2009qs} the rescaling of the $B$ couplings is performed but the hat symbol is not explicitly introduced.} $\hat B_{bs}^L \equiv B_{bs}^L (g_2 m_Z/g_1 m_Z^\prime)$. The reason for replacing $B_{bs}^L$ with the $m_{Z^\prime}$ dependent coupling $\hat B_{bs}^L $ is that the latter is directly constrained by data on $B_s$ mixing and $B\to (\pi,\psi,\phi,\eta^\prime.\rho,\omega,f^0)K_S$ decays: the authors of Ref.~\cite{Barger:2009qs} find that these constraints result into the upper bound $|\hat B_{bs}^L| \lesssim 2 \times 10^{-3}$. From the inspection of Eqs.~(\ref{eq:c9zprime}) and (\ref{eq:c10zprime}) it is clear that the exchange of a $Z^\prime$ generates only sizable contributions to $C_{10}$. In fact, as a direct consequence of our assumptions of identical $Z$ and $Z^\prime$ couplings to leptons we have $\delta C_9/\delta C_{10}  =  4 s_W^2 -1 \simeq -0.075$. For this reason we expect large contributions to ${\cal R}[4]$ only for (large) non-zero values of the $\hat B_{bs}^L$ phase.  

We present the results of our numerical analysis in Fig.~\ref{fig:zprime}, where we show the dependence of ${\cal R}[4]$ on the modulus and phase of $\hat B_{bs}^L$ for $m_{Z^\prime} = 1 \; {\rm TeV}$. Clearly very large effects are possible and a measurement of ${\cal R}[4]$, together with CP violation in $B_s$ mixing and $b\to s q\bar q$ modes can provide a smoking gun for these models.

\section{Conclusions}
\label{sec:conclusions}
In this paper we introduce a new observable in exclusive $B\to K^* \ell^+ \ell^-$ and $B_s\to \phi \ell^+ \ell^-$ decays: the ratio ${\cal R}[4]$ of the integrated forward--backward asymmetry in the $[4,6]\gev^2$ and $[1,4]\gev^2$ bins. The separation between the two bins roughly coincides with the position of the zero of the FB asymmetry spectrum (calculated in the SM at NLO). 

We have shown that the bulk of the theoretical uncertainty on ${\cal R}[4]$ is due to perturbative scale uncertainties and that the form factor dependence drops out almost completely. This ratio is therefore extremely interesting because its SM prediction Eq.~(\ref{ratioSM}) can be systematically improved without requiring any non-perturbative input. The position of the zero of the spectrum shares these positive features and can be determined with even better theoretical accuracy, but is much harder to measure. 

A very interesting feature of ${\cal R}[4]$ is the strong dependence on the CP violating phases (especially those appearing in the Wilson coefficient $C_{10}$). This dependence arises because of the presence of strong phases in the $b\to s\ell^+\ell^-$ matrix elements of current--current and QCD penguin operators (i.e. $O_{1-6}$). The latter can be determined perturbatively using QCD factorization (or SCET). 

We performed a model independent study of new physics contributions to the coefficients $C_{7,9,10}$ and specialized the analysis to several concrete new physics models. We find that, after imposing the constraints from experimental data on radiative and semileptonic rare decays, there are two scenarios in which large effects are possible.
\begin{list}{\labelitemi}{\leftmargin=1em}
\item Models with modest (large) contributions to the coefficients $C_7$ ($C_9$). As a representative of this class we considered an MSSM with non--vanishing $(\delta_{23}^D)_{LR}$ mass insertion. In this case, large effects on ${\cal R}[4]$ and on the location of the zero are correlated. Moreover, we do not find a very strong dependence on the new physics phases of $\delta C_{7,9}$. 
\item Models with large and complex contributions to $C_{10}$. We considered the inclusion of a sequential 4th generation to the SM (${\rm SM}4$) and a $U(1)^\prime$ model with family--dependent couplings (i.e. tree-level flavor changing $Z^\prime$ interactions). In ${\rm SM}4$ we find large contributions to $C_{10}$ because of the loop--function dependence on the $t^\prime$ mass; in the family dependent $U(1)^\prime$ model this behavior is a result of our assumption of universal $Z$ and $Z^\prime$ couplings to leptons. In these models we find very large effects on ${\cal R}[4]$ driven by the phase of $\delta C_{10}$. At the same time there is almost no change in the position of the zero of the spectrum because new physics contributions to $C_{10}$ do not impact this observable at all.
\end{list}
In conclusion, we believe that this new observable should be promptly included in future experimental studies of $B\to K^* \ell\ell$ decays because it allows for an easier access (i.e. less luminosity is required) to the new physics probed by the zero of the spectrum and, at the same time, is sensitive to the quite elusive new physics phase in $C_{10}$.

\section*{Acknowledgments}
This research was supported in part by the U.S. DOE contract No.DE-AC02-98CH10886(BNL).

\end{document}